\documentclass[journal]{IEEEtran}
\usepackage{graphicx}
\usepackage{multirow}
\usepackage[table,xcdraw,dvipsnames]{xcolor}
\usepackage{color}
\usepackage{xcolor,colortbl}
\usepackage{tikz}
\def\checkmark{\tikz\fill[scale=0.4](0,.35) -- (.25,0) -- (1,.7) -- (.25,.15) -- cycle;}
\usepackage{pifont}
\usepackage[activate]{microtype}
\newcommand{\xmark}{\text{\ding{55}}}
\usepackage{authblk}
\usepackage{booktabs}% http://ctan.org/pkg/booktabs
\newcommand{\tabitem}{~~\llap{\textbullet}~~}
\usepackage{url}
\usepackage[dvipsnames]{xcolor}
\usepackage{soul}

\usepackage[separate-uncertainty = true,multi-part-units = repeat]{siunitx}
\usepackage{array,amsfonts,amsmath}
\usepackage{caption}
\usepackage{verbatim}
\ifCLASSOPTIONcompsoc
  \usepackage[nocompress]{cite}
\else
  \usepackage{cite}
\fi
\ifCLASSINFOpdf
\else
\fi

\begin{document}

%\title{A Survey on Deep Reinforcement Learning for Audio Processing}
\title{A Survey on Deep Reinforcement Learning\\for Audio-Based Applications}

\author[1]{Siddique Latif\thanks{Email: siddique.latif@usq.edu.au}}
\author[2]{Heriberto Cuay\'ahuitl}
\author[3]{Farrukh Pervez}
\author[4]{Fahad Shamshad}
\author[5]{Hafiz Shehbaz Ali}
\author[6]{Erik Cambria}
%\author[6]{Soujanya Poria}

\affil[1]{University of Southern Queensland, Australia}
\affil[2]{University of Lincoln, United Kingdom}
\affil[3]{National University of Science and Technology, Pakistan}
\affil[4]{Information Technology University, Pakistan}
\affil[5]{EmulationAI}
\affil[6]{Nanyang Technological University, Singapore}
%\affil[6]{Singapore University of Technology and Design}

\maketitle

\begin{abstract}
Deep reinforcement learning (DRL) is poised to revolutionise the field of artificial intelligence (AI) by endowing autonomous systems with high levels of understanding of the real world. Currently, deep learning (DL) is enabling DRL to effectively solve various intractable problems %. This enables DRL applications 
in various fields. %including audio domain. 
Most importantly, DRL algorithms are also being employed in audio signal processing to learn directly from speech, music and other sound signals in order to create audio-based autonomous systems that have many promising application in the real world. In this article, we conduct a comprehensive survey on the progress of DRL in the audio domain by bringing together the research studies across different speech and music related areas. We begin with an introduction to the general field of DL and reinforcement learning (RL), then progress to the main DRL methods and their applications in the audio domain. %: value-based, policy-based and model-based. %The main contribution of this paper is an up-to-date and comprehensive survey on DRL for audio-based application. % in the fields of speech, music, environmental sound signal processing, and dialogue and robotic agents. 
%In addition, we also present challenges faced by audio-based DRL agents and highlight open areas for future research and investigation.
We conclude by presenting challenges faced by audio-based DRL agents and highlighting open areas for future research and investigation.
\end{abstract}


\IEEEpeerreviewmaketitle

\section{Introduction}
\label{sec:introduction}
Artificial intelligence (AI) has gained widespread attention in many areas of life, especially in audio signal processing. Audio processing covers many diverse fields including speech, music and environmental sound processing. In all these areas, AI techniques are playing crucial roles in the design of audio-based intelligent systems~\cite{purwins2019deep}. One of the prime goals of AI is to create fully autonomous audio-based intelligent systems or agents that can learn optimal behaviours by listening or interacting with their environments and improving their behaviour over time through trial and error. Designing of such autonomous systems has been a long-standing problem, ranging from robots that can react to the changes in their environment, to purely software-based agents that can interact with humans using natural language and multimedia. Reinforcement learning (RL)~\cite{sutton1998introduction} represents a principled mathematical framework of such experience-driven learning. Although RL had some successes in the past~\cite{kohl2004policy,ng2006autonomous,singh2002optimizing}, however, previous methods were inherently limited to low-dimensional problems due to lack of scalability. Moreover, RL also has issues of memory,  computational and sample complexity---in the case of learning algorithms~\cite{strehl2006pac}. Recently, deep learning (DL) models have risen as new tools with powerful function approximation and representation learning properties to solve these issues. 

The advent of DL has had a significant impact on many areas of machine learning (ML) by dramatically improving the state-of-the-art in image processing tasks such as object detection and image classification. Deep models such as deep neural networks (DNNs)~\cite{hinton2012deep,mohamed2009deep}, convolutional neural networks (CNNs)~\cite{lecun1989backpropagation}, and long short-term memory (LSTM) networks~\cite{hochreiter1997long} have also enabled many practical applications by outperforming traditional methods in audio signal processing. Given that DL has also accelerated RL's progress with the use of DL algorithms within RL, it has given rise to the field of deep reinforcement learning (DRL).

DRL embraces the advancements in DL to establish the learning processes, performance and speed of RL algorithms. This enables RL to operate in high-dimensional state and action spaces to solve complex problems that were previously difficult to solve. As a result, DRL has been adopted to solve many problems. Inspired by previous works such as~\cite{LangeRV12}, two outstanding works kick-started the revolution in DRL. The first was the development of an algorithm that could learn to play Atari 2600 video games directly from image pixels at a superhuman level~\cite{mnih2015human}. The second success was design of the hybrid DRL system, AlphaGo, which defeated a human world champion in the game of Go~\cite{silver2016mastering}. %, and the historic achievement of IBM's Deep Blue in chess~\cite{campbell2002deep} and IBM's Watson DeepQA system that beat the best human Jeopardy! players~\cite{ferrucci2010building}. 
In addition to playing games, DRL has also been applied to a wide range of problems including robotics to control policies~\cite{levine2016end}; generalisable agents in complex environments with meta-learning~\cite{duan2016rl,wang2016learning}; indoor navigation~\cite{zhu2017target}, and many more~\cite{8103164}. In particular, DRL is also gaining increased interest in audio signal processing.

\begin{table*}[!ht]
\centering
\caption{Comparison of our paper with that of the existing surveys.}
\begin{tabular}{|l|l|l|l|l|}
\hline
 & \multicolumn{3}{c|}{\textbf{Focus}}   &                                               \\ \cline{2-4}
 
\multirow{-2}{*}{\bf Reference}                        & \begin{tabular}[c]{@{}l@{}}\bf Deep\\ \bf Reinforcement\\ \bf Learning\end{tabular} & \begin{tabular}[c]{@{}l@{}}\bf Audio\\ \bf Applications\end{tabular} & \begin{tabular}[c]{@{}l@{}}\bf Other\\ \bf Applications\end{tabular} &
\multirow{-2}{*}{\bf Details}                     \\ \hline \hline

\begin{tabular}[c]{@{}l@{}}Arulkumaran et al.~\cite{8103164}\\2017\end{tabular} & \begin{tabular}[c]{@{}l@{}}\checkmark{}\end{tabular}                           & \begin{tabular}[c]{@{}l@{}}\xmark{}\end{tabular}                & \begin{tabular}[c]{@{}l@{}}\xmark{}\end{tabular}     
& \begin{tabular}[c]{@{}l@{}}This paper presents a brief overview of recent developments in\\DRL algorithms and highlights the benefits of DRL and several \\current areas of research.  \end{tabular} \\ \hline
\begin{tabular}[c]{@{}l@{}}Yuxi Li \cite{li2017deep}\\2017\end{tabular}  & \begin{tabular}[c]{@{}l@{}}\checkmark{}\end{tabular}  
& \begin{tabular}[c]{@{}l@{}}\xmark{}\end{tabular} & \begin{tabular}[c]{@{}l@{}}\checkmark{}\end{tabular}
& \begin{tabular}[c]{@{}l@{}}This paper presents a generalised overview of recent exciting \\achievements of DRL and discus core elements and mechanisms.\\ It also discusses various fields where DRL can be applied. \end{tabular} \\ \hline

\begin{tabular}[c]{@{}l@{}}Luong et al. \cite{luong2019applications}\\2019\end{tabular}  & \begin{tabular}[c]{@{}l@{}}\checkmark{}\end{tabular}                           & \begin{tabular}[c]{@{}l@{}}\xmark{}\end{tabular}  & \begin{tabular}[c]{@{}l@{}}communications\\ and networking\end{tabular} 
& \begin{tabular}[c]{@{}l@{}}This paper presents a comprehensive literature review on the\\ applications of DRL in communications and networking, highlights\\ challenges, and discusses open issues and future directions. \end{tabular} \\ \hline

\begin{tabular}[c]{@{}l@{}}Kiran et al. \cite{kiran2020deep}\\2020\end{tabular}  & \begin{tabular}[c]{@{}l@{}}\checkmark{}\end{tabular}  
& \begin{tabular}[c]{@{}l@{}}\xmark{}\end{tabular}  & \begin{tabular}[c]{@{}l@{}}autonomous \\driving\end{tabular}               
& \begin{tabular}[c]{@{}l@{}}This paper summarises DRL algorithms and autonomous driving, \\where (D)RL methods
have been employed.It also highlights key\\challenges towards real-world deployment of autonomous cars. 

\end{tabular} \\ \hline

\begin{tabular}[c]{@{}l@{}}Haydari et al. \cite{haydari2020deep}\\2020\end{tabular}  & \begin{tabular}[c]{@{}l@{}}\checkmark{}\end{tabular}                           & \begin{tabular}[c]{@{}l@{}}\xmark{}\end{tabular}  & \begin{tabular}[c]{@{}l@{}}transportation\\ systems\end{tabular}                & \begin{tabular}[c]{@{}l@{}}This paper summarises existing works in the field of transportation,\\ and discusses the challenges and open questions regarding DRL\\ in transportation systems.\end{tabular} \\ \hline
\rowcolor[HTML]{EFEFEF} 
Ours (2020)                                           & \begin{tabular}[c]{@{}l@{}}\checkmark{}\end{tabular}                           & \begin{tabular}[c]{@{}l@{}}\checkmark{}\end{tabular}                & \begin{tabular}[c]{@{}l@{}}\xmark{}\end{tabular}                & \begin{tabular}[c]{@{}l@{}}We present a comprehensive review focused on DRL applications\\ in the audio domain, highlight existing challenges that hinder the \\progress of DRL in audio, and discuss pointers for future research.\end{tabular} \\ \hline
\end{tabular}
\label{Table:compa}
\end{table*}

In audio processing, DRL has been recently used as an emerging tool to address various problems and challenges in automatic speech recognition (ASR), spoken dialogue systems (SDSs), speech emotion recognition (SER), audio enhancement, music generation, and audio-driven controlled robotics. In this work, we therefore focus on covering the advancements in audio processing by DRL. Although there are multiple survey articles on DRL. For instance, Arulkumaran et al.~\cite{8103164} presented a brief survey on DRL by covering seminal and recent developments in DRL---including innovative ways in which DNNs can be used to develop autonomous agents. Similarly, in~\cite{li2017deep}, authors attempted to provide comprehensive details on DRL and cover its applications in various areas to highlight advances and challenges.  Other relevant works include applications of DRL in communications and networking~\cite{luong2019applications}, human-level agents~\cite{nguyen2017system}, and autonomous driving~\cite{sallab2017deep}. None of these articles has focused on DRL applications in audio processing as highlighted in Table \ref{Table:compa}. This paper aims to fill this gap by presenting an up-to-date literature review on DRL studies in the audio domain, discussing challenges that hinder the progress of DRL in audio, and pointing out future research areas. We hope this paper will help researchers and scientists interested in DRL for audio-driven applications.

This paper is organised as follows. A concise background of DL and RL is provided in Section~\ref{back}, followed by an overview of recent DRL algorithms in Section~\ref{sec:DRL}. With those foundations, Section~\ref{survey} covers recent DRL works in domains such as speech, music, and environmental sound processing; and their challenges are discussed in Section~\ref{challenges}. Section \ref{summary} summaries this review and highlights the future pointers for audio-based DRL research and Section~\ref{conclu} concludes the paper.

\section{Background}
\label{back}

\subsection{Deep Learning (DL)}
\begin{figure*}[!ht]%[!ht]
\centering
\includegraphics[width=0.95\textwidth]{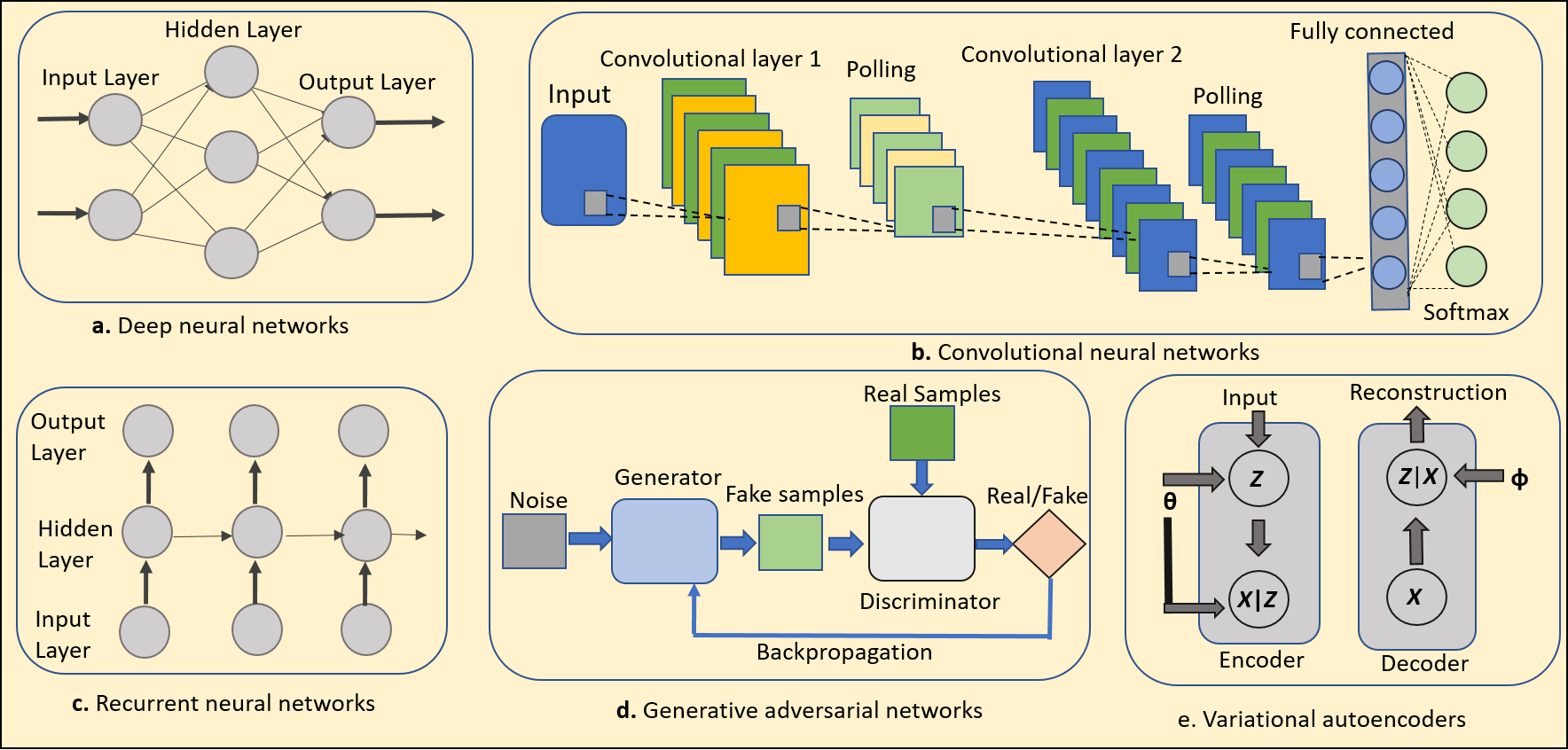}
\caption{Graphical illustration of different DL architectures.}
\label{fig:models}
\end{figure*}

DNNs have been shown to produce state-of-the-art results in audio and speech processing due to their ability to distil compact and robust representations from large amounts of data. The first major milestone was significantly increasing the accuracy of large-scale automatic speech recognition based on the use of fully connected DNNs and deep autoencoders around 2010~\cite{hinton2012deep}. It focuses on the use of artificial neural networks, which consists of multiple nonlinear modules arranged hierarchically in layers to automatically discover suitable representations or features from raw data for specific tasks. These non-linearities allow DNNs to learn complicated manifolds in speech and audio datasets. Below we discuss different DL architectures, which are illustrated in Figure~\ref{fig:models}.

\textbf{Convolutional neural networks (CNNs)} are a kind of feedforward neural networks that have been specifically designed for processing  data having grid-like topologies such as images~\cite{krizhevsky2012imagenet}. Recently, CNN's have shown state-of-the-art performance in various image processing tasks, including segmentation, detection, and classification, among others \cite{KhanSZQ20}. In contrast to DNNs, CNNs limit the number of parameters and memory requirements dramatically by leveraging on two key concepts: \textit{local receptive fields} and \textit{shared weights}. They often consist of a series of convolutional layers interleaved with pooling layers, followed by one or more dense layers. For sequence labelling, the dense layers can be omitted to obtain a fully-convolutional network (FCN). FCNs have been extended with domain adaptation for increased robustness~\cite{TzengEtAl2017}. Recently, CNN models have been extensively studied for a variety of audio processing tasks including music onset detection~\cite{6854953}, speech enhancement~\cite{mamun2019convolutional}, ASR~\cite{abdel2014convolutional}, etc. However, raw audio waveform with high sample rates might have problems with limited receptive fields of CNNs, which can result in deteriorated performance. To handle this performance issue, dilated convolution layers can be used in order to extend the receptive field by inserting zeros between their filter coefficients~\cite{chang2018temporal, chen2019environmental}. 

\textbf{Recurrent neural networks (RNNs)} follow a different approach for modelling sequential data~\cite{Lipton15}. %They compute the output of the current time step from both the input at that step and their hidden state at the previous step. This allows RNNs to inherently learns the temporal structures in the input data. %and allows the receptive field to extend indefinitely into the past.
They introduce recurrent connections to enable parameters to be shared across time, which make them very powerful in learning temporal structures from the input sequences (e.g., audio, video). They have demonstrated their superiority over traditional HMM-based systems in a variety of speech and audio processing tasks~\cite{latif2020speech}. Due to these abilities, RNNs, especially long-short term memory (LSTM)~\cite{hochreiter1997long} and gated recurrent unit (GRU)~\cite{cho2014learning} networks, have had an enormous impact in the speech community, and they are incorporated in state-of-the-art audio-based systems. Recently, these RNN models have been extended to include information in the frequency domain besides temporal information in the form of Frequency-LSTMs \cite{li2015lstm} and Time-Frequency LSTMs \cite{sainath2016modeling}. In order to benefit from both neural architectures, CNNs and RNNs can be combined into a single network with convolutional layers followed by recurrent layers, often referred to as convolutional recurrent neural networks (CRNN). Related works combining CNNs and RNNs have been presented in ASR~\cite{qian2016very}, SER \cite{latifdeep}, music classification~\cite{ghosal2018music}, and other audio related applications~\cite{latif2020speech}.

\textbf{Sequence-to-sequence (Seq2Seq) models} were motivated due to problems requiring sequences of unknown lengths~\cite{sutskever2014sequence}. Although they were initially applied to machine translation, they can be applied to many different applications involving sequence modelling. In a Seq2Seq model, while one RNN reads the inputs in order to generate a vector representation (the {\it encoder}), another RNN inherits those learnt features to generate the outputs (the {\it decoder}). The neural architectures can be single or multilayer, unidirectional or bidirectional~\cite{SchusterP97}, and they can combine multiple architectures \cite{Lipton15,abs-1801-01078} using end-to-end learning by optimising a joint objective instead of independent ones. Seq2Seq models have been gaining much popularity in the speech community due to their capability of transducing input to output sequences. DL frameworks are particularly suitable for this direct translation task due to their large model capacity and their capability to train in an end-to-end manner---to directly map the input signals to the target sequences~\cite{zhang2017very,lu2016training,liu2019new}. Various Seq2Seq models have been explored in the speech, audio and language processing literature including Recurrent Neural Network Transducer (RNNT)~\cite{graves2012sequence}, Monotonic Alignments~\cite{raffel2017online}, Listen, Attend and Spell (LAS)~\cite{chan2015listen}, Neural Transducer~\cite{jaitly2016online}, Recurrent Neural Aligner (RNA)~\cite{raffel2017online}, and Transformer Networks~\cite{PhamNN0W19}, among others.

\textbf{Generative Models} have been attaining much interest in the three types of generative models: Generative Adversarial Networks (GANs)~\cite{goodfellow2014generative}, Variational Autoencoders (VAEs)~\cite{kingma2013auto},  and autoregressive models~\cite{shannon2012autoregressive}. This type of models are powerful enough to learn the underlying distribution of speech datasets and have been extensively investigated in the speech and audio processing scientific community. Specifically, in the case of GANs and VAEs, audio signal is often synthesised from a low-dimensional representation, from which it needs to by upsampled (e.g., through linear interpolation or the nearest neighbour) to the high-resolution signal~\cite{hsu2017learning,ma2019m3d}. %Artefacts,induced by the different layer resolutions, can be ameliorated through random phase perturbation in different layers.
Therefore, VAEs and GANs have been extensively explored for synthesising speech or to augment the training material by generating features \cite{latif2020deep} or speech itself. In the autoregressive approach, the new samples are synthesised iteratively---based on an infinitely long context of previous samples via RNNs (for example, using LSTM or GRU networks)---but at the cost of expensive computation during training~\cite{wang2018autoregressive}.

\subsection{Reinforcement learning}
\begin{figure}[!ht]
\centering
\includegraphics[width=0.38\textwidth]{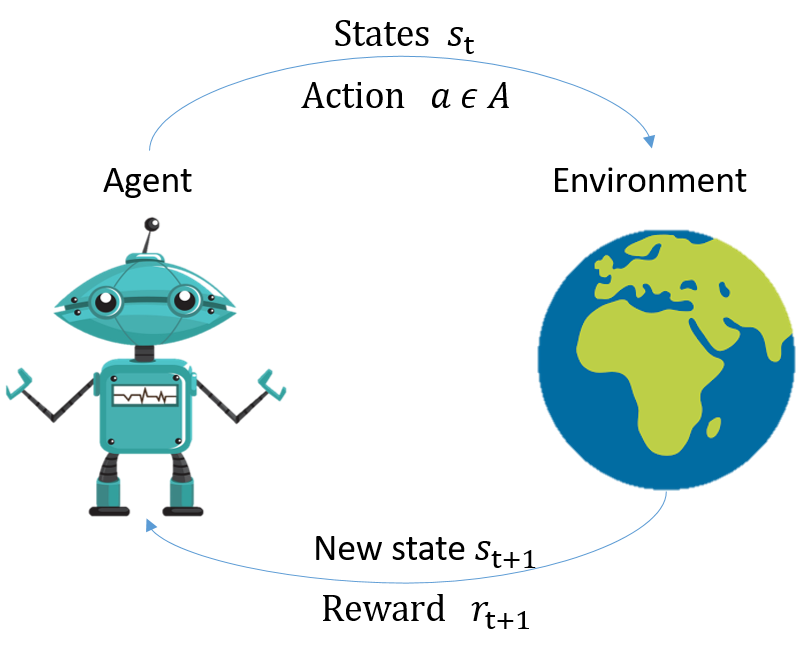}
\caption{Basic RL setting.}
\label{fig:RL}
\end{figure}

Reinforcement learning (RL) is a popular paradigm of ML, which involves agents to learn their behaviour by trial and error~\cite{sutton1998introduction}. RL agents aim to learn sequential decision-making by successfully interacting with the environment where they operate. At time $t$ (0 at the beginning of the interaction, $T$ at the end of an episodic interaction, or $\infty$ in the case of non-episodic tasks), an RL agent in state $s_{t}$ takes an action $a\in A$, transits to a new state $s_{t+1}$, and receives reward $r_{t+1}$ for having chosen action $a$. This process---repeated iteratively---is illustrated in Figure~\ref{fig:RL}.
\\
An RL agent aims to learn the best sequence of actions, known as {\it policy}, to obtain the highest overall cumulative reward in the task (or set of tasks) that is being trained on. While it can choose any action from a set of available actions, the set of actions that an agent takes from start to finish is called an {\it episode}. A Markov decision process (MDP)~\cite{bellman1966dynamic} can be used to capture the episodic dynamics of an RL problem. An MDP can be represented using the tuple ($S$, $A$, $\gamma$, $P$, $R$). The decision-maker or agent chooses an action $a \in A$ in state $s \in S$ at time $t$ according to its policy $\pi (a_{t}|s_{t})$---which determines the agent's way of behaving. The probability of moving to the next state $s_{t+1} \in S$ is given by the state transition function $P(s_{t+1}|s_{t}, a_{t})$. The environment produces a reward $R(s_t, a_t, s_{t+1})$ based on the action taken by the agent at time $t$. This process continues until the maximum time step or the agent reaches a terminal state. 
The objective is to maximise the expected discounted cumulative reward, which is given by:

\begin{equation}%\nonumber
    E_{\pi}[R_{t}]=E_{\pi}\Big[\sum_{i=0}^{\infty}\gamma^{i}r_{t+i}\Big]
\end{equation}
where $\gamma$ $\in$ [0,1] is a discount factor used to specify that rewards in the distant future are less valuable than in the nearer future. 
%to penalise distant rewards. 
While an RL agent may only learn its policy, it may also learn (online or offline) the transition and reward functions.

\section{Deep Reinforcement Learning}
\label{sec:DRL}
\begin{figure*}[th]%[!ht]
\centering
\includegraphics[width=0.8\textwidth]{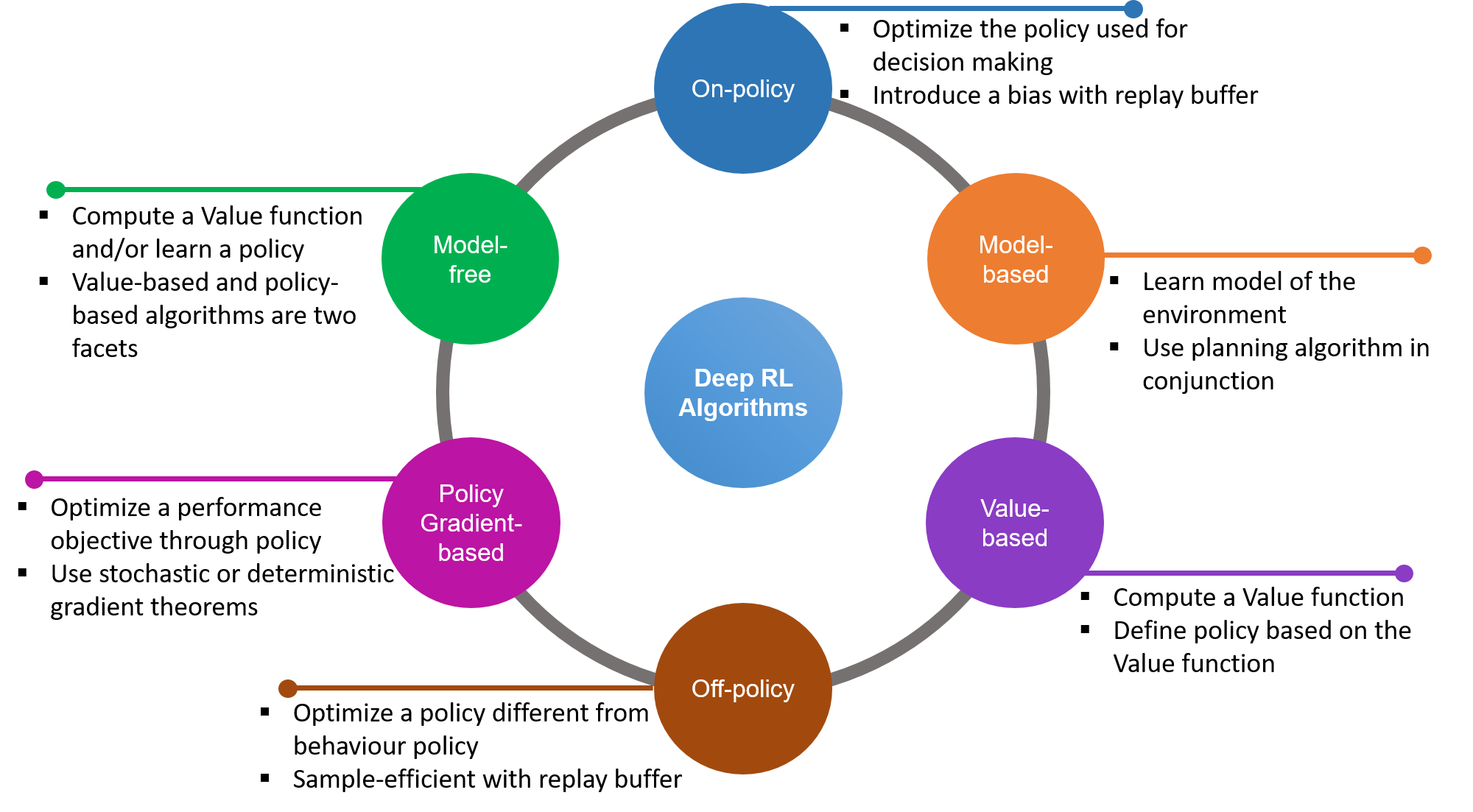}
\caption{Classification of DRL algorithms.}
\label{fig:DRL_C}
\end{figure*}

Deep reinforcement learning (DRL) combines conventional RL with DL to overcome the limitations of RL in complex environments with large state spaces or high computation requirements. DRL employs DNNs to estimate value, policy or model that are learnt through the storage of state-action pairs in conventional RL~\cite{li2017deep}. Deep RL algorithms can be classified along several dimensions. For instance, on-policy vs off-policy, model-free vs model-based, value-based vs policy-based  DRL algorithms, among others. The salient features of various key categories of DRL algorithms are presented and depicted in Figure \ref{fig:DRL_C}. Interested readers are referred to~\cite{li2017deep} for more details on these algorithms. This section focuses on popular DRL algorithms employed in audio-based applications in three main categories: (i) value-based DRL, (ii) policy gradient-based  DRL and (iii) model-based DRL.  

\subsection{Value-Based DRL}
One of the most famous value-based DRL algorithms is Deep Q-network (DQN), introduced by Mnih et al.~\cite{mnih2015human}, that learns directly from high-dimensional inputs. It employs convolution neural networks (CNNs) to estimate a value function $Q(s, a)$, which is subsequently used to define a policy. DQN enhances the stability of the learning process using the concept of target Q-network along with experience replay. The loss function computed by DQN at $i^{th}$ iteration is given by 
\begin{equation}%\nonumber
\begin{aligned}
    L_{i}(\theta_{i})=\mathbb{E}_{s,a\sim p(.)}[(y_{i}-Q(s,a;\theta_{i}))^2],\\
    \text{where} \quad y_{i}=\mathbb{E}_{s\prime \sim s}[r+\gamma \underset{a\prime}{\text{max}} Q(s\prime, a\prime;\theta_{i-1}|{s,a}].
\end{aligned}
\end{equation}
Although DQN, since inception, has rendered super-human performance in Atari games, it is based on a single max operator, given in (2), for selection as well evaluation of an action. Thus, the selection of an overestimated action may lead to over-optimistic action value estimates that induces an upward bias. Double DQN (DDQN)~\cite{van2016deep} eliminates this positive bias by introducing two decoupled estimators: one for the selection of an action, and one for the evaluation of an action. Schaul et al. in~\cite{schaul2015prioritized} show that the performance of DQN and DDQN is enhanced considerably if significant experience transitions are prioritised and replayed more frequently. Wang et al.~\cite{wang2016dueling} present a duelling network architecture (DNA) to estimate a value function $V(s)$ and associated advantage function $A(s, a)$ separately, and then combine them to get action-value function $Q(s, a)$. Results prove that DQN and DDQN having DNA and prioritised experience replay can lead to improved performance. 

Unlike the aforementioned DQN algorithms that focus on the expected return, distributional DQN~\cite{bellemare2017distributional} aims to learn the full distribution of the value in order to have additional information about rewards. Despite both DQN and distributional DQN focusing on maximising the expected return, the latter comparatively results in performant learning. Will et al.~\cite{dabney2018distributional} propose distributional DQN with quantile regression (QR-DQN) to explicitly model the distribution of the value function. Results prove that QR-DQN successfully bridges the gap between theoretic and algorithmic results. Implicit Quantile Networks (IQN)~\cite{dabney2018implicit}, an extension to QR-DQN, estimate quantile regression by learning the full quantile function instead of focusing on a discrete number of quantiles. IQN also provides flexibility regarding its training with the required number of samples per update, ranging from one sample to a maximum computationally allowed. IQN has shown to outperform QR-DQN comprehensively in the Atari domain.

The astounding success of DQN to learn rich representations is highly attributed to DNNs, while batch algorithms prove to have better stability and data efficiency (requiring less tuning of hyperparameters). Authors in~\cite{levine2017shallow} propose a hybrid approach named as Least Squares DQN (LS-DQN) that exploits the advantages of both DQN and batch algorithms. Deep Q-learning from demonstrations (DQfD)~\cite{hester2018deep} leverages human demonstrations to learn at an accelerated rate from the start. Deep Quality-Value (DQV)~\cite{sabatelli2018deep} is a novel temporal-difference-based algorithm that trains the Value network initially, and subsequently uses it to train a Quality-value neural network for estimating a value function. Results in the Atari domain indicate that DQV outperforms DQN as well as DDQN. Authors in~\cite{arjona2019rudder} propose RUDDER (Return Decomposition for Delayed Rewards), which encompasses reward redistribution and return decomposition for Markov decision processes (MDPs) with delayed rewards. Pohlen et al.~\cite{pohlen2018observe} employ a transformed Bellman operator along with human demonstrations in the proposed algorithm Ape-X DQfD to attain human-level performance over a wide range of games. Results prove that the proposed algorithm achieves average-human performance in 40 out of 42 Atari games with the same set of hyperparameters. Schulman et al. in~\cite{schulman2017equivalence} study the connection between Q-learning and policy gradient methods. They show that soft Q-learning (an entropy-regularised version of Q-learning) is equivalent to policy gradient methods and that they perform as well (if not better) than standard variants.
%Authors show equivalence between soft Q-learning (entropy-regularised version of Q-learning) and policy gradient method concerning expected gradients holds relevance practically.                 

%In the audio domain, DNNs with memory units like LSTM has shown great performance, especially for speech modeling. 
Previous studies have also attempted to incorporate the memory element into DRL algorithms. For instance, the deep recurrent Q-network (DRQN) approach introduced by~\cite{hausknecht2015deep} was able to successfully integrate information through time, which performed well on standard Atari games. A further improvement was made by introducing an attention mechanism to DQN, resulting in a deep recurrent Q-network (DARQN)~\cite{sorokin2015deep}. This allows DARQN to focus on a specific part of the input and achieve better performance compared to DQN and DRQN on games. Some other studies~\cite{oh2016control,parisotto2017neural} have also proposed methods to incorporate memory into DRL, but this area remains to be investigated further.
%; however, they are not focused on the audio signal. Although attention-based recurrent networks have been evaluated in audio domain  but it will be interesting to design DRL algorithms that can focus on the specific part of the audio signal and can learn contexts from these parts.    

\subsection{Policy Gradient-Based DRL}

Policy gradient-based DRL algorithms aim to learn an optimal policy that maximises performance objectives, such as  expected cumulative reward. This class of algorithms make use of gradient theorems to reach optimal policy parameters. Policy gradient typically requires the estimation of a value function based on the current policy. This may be accomplished using the actor-critic architecture, where the {\it actor} represents the policy and the {\it critic} refers to value function estimate~\cite{konda1999actor}. Mnih et al.~\cite{mnih2016asynchronous} show that asynchronous execution of multiple parallel agents on standard CPU-based hardware leads to time-efficient and resource-efficient learning. The proposed asynchronous version of actor-critic, asynchronous advantage actor-critic (A3C) exhibit remarkable learning in both 2D and 3D games with action spaces in discrete as well as continuous domains. Authors in~\cite{babaeizadeh2017reinforcement} propose a hybrid CPU/GPU-based A3C ---named as GA3C --- showing significantly higher speeds as compared to its CPU-based counterpart.

Asynchronous actor-critic algorithms, including A3C and GA3C, may suffer from inconsistent and asynchronous parameter updates. A novel framework for asynchronous algorithms is proposed in~\cite{alfredo2017efficient} to leverage parallelisation while providing synchronous parameters updates. Authors show that the proposed parallel advantage actor-critic (PAAC) algorithm enables true on-policy learning in addition to faster convergence. Authors in~\cite{o2016pgq} propose a hybrid policy-gradient-and-Q-learning (PGQL) algorithm that combines on-policy policy gradient with off-policy Q-learning. Results demonstrate PGQL's superior performance on Atari games as compared to both A3C and Q-learning. Munos et al.~\cite{munos2016safe} propose a novel algorithm by bringing together three off-policy algorithms: Instance Sampling (IS), Q($\lambda$), and Tree-Backup TB($\lambda$). This algorithm --- called Retrace($\lambda$) --- alleviates the weaknesses of all three algorithms (IS has low variance, Q($\lambda$) is not safe, and TB($\lambda$) is inefficient) and promises safety, efficiency and guaranteed convergence. Reactor (Retrace-Actor)~\cite{gruslys2017reactor} is a Retrace-based actor-critic agent architecture that combines time efficiency of asynchronous algorithms with sample efficiency of off-policy experience replay-based algorithms. Results in the Atari domain indicate that the proposed algorithm performs comparably with state-of-the-art algorithms while yielding substantial gains in terms of training time. The importance of weighted actor-learner architecture (IMPALA)~\cite{espeholt2018impala} is a scalable distributed agent that is capable of handling multiple tasks with a single set of parameters. Results show that IMPALA outperforms A3C baselines in a diverse multi-task environment. 

Schulman et al.~\cite{schulman2015trust} propose a robust and scalable trust region policy optimisation (TRPO) algorithm for optimising stochastic control policies. TRPO promises guaranteed monotonic improvement regarding the optimisation of nonlinear and complex policies having an inundated number of parameters. This learning algorithm makes use of a fixed KL divergence constraint rather than a fixed penalty coefficient, and outperforms a number of gradient-free and policy-gradient methods over a wide variety of tasks.~\cite{schulman2017proximal} introduce proximal policy optimisation (PPO), which aims to be as reliable and stable as TRPO but relatively better in terms of implementation and sample complexity.

% \usepackage{multirow}
\begin{table*}[!ht]
\scriptsize
\caption{Summary of DRL algorithms.}
\label{Summary of DRL algorithms}
\centering
\begin{tabular}{|l|l|l|l|}
\hline
\textbf{DRL algorithms} & \textbf{Approach}                                                                                                                                                                                             & \textbf{Details}                                                                                                                                                                                      & \textbf{\begin{tabular}[c]{@{}l@{}}off-policy/\\ on policy\end{tabular}} \\ \hline \hline
\multicolumn{4}{|c|}{\textbf{Value-based DRL}}                                                                                                                                                                                                                                                                                                                                                                                                                                                                            \\ \hline\hline
DQN {[}2{]}                 & Target Q-network, experience replay                                                                                                                                                                           & \begin{tabular}[c]{@{}l@{}}\tabitem Learns directly from high dimensional visual inputs\\ \tabitem Stabilises learning process with target Q-network\\ \tabitem Experience replay to avoid divergence in parameters\end{tabular} & \multirow{13}{*}{off-policy}                                             \\ \cline{1-3}
DDQN {[}3{]}                & Double Q-learning                                                                                                                                                                                             & \begin{tabular}[c]{@{}l@{}}Decoupled estimators for the selection and evaluation of an action\end{tabular}                                                                                    &                                                                         \\ \cline{1-3}
Prioritised DQN {[}4{]}     & Prioritised experience replay                                                                                                                                                                                 & \begin{tabular}[c]{@{}l@{}}Significant experience transitions are prioritised and replayed\\  frequently thus leading to efficient learning\end{tabular}                                              &                                                                         \\ \cline{1-3}
DNA {[}5{]}                 & Duelling neural network architecture                                                                                                                                                                           & \begin{tabular}[c]{@{}l@{}}Estimates a value function and associated advantage function and combine\\ them to get a value function with faster convergence than Q-learning\end{tabular}         &                                                                         \\ \cline{1-3}
Distributional DQN {[}6{]}  & \begin{tabular}[c]{@{}l@{}}Learns distribution of cumulative returns\\  using distributional Bellman equation\end{tabular}                                                                                    & \begin{tabular}[c]{@{}l@{}}\tabitem Leads to performant learning than DQN\\\tabitem Possibility   to implement risk-aware behaviour\end{tabular}                                                                       &                                                                         \\ \cline{1-3}
QR-DQN {[}7{]}              & \begin{tabular}[c]{@{}l@{}}Distributional DQN with quantile regression\end{tabular}                                                                                                                        & Bridges gap between theoretical and algorithmic results                                                                                                                                               &                                                                         \\ \cline{1-3}
IQN {[}8{]}                 & \begin{tabular}[c]{@{}l@{}}Extends QR-DQN with a full quantile function\end{tabular}                                                                                                                    & \begin{tabular}[c]{@{}l@{}}Provides flexibility regarding number of samples required for training\end{tabular}                                                                                     &                                                                         \\ \cline{1-3}
LS-DQN {[}9{]}              & \begin{tabular}[c]{@{}l@{}}A hybrid approach combining DQN with\\ least-squares method\end{tabular}                                                                                                           & \begin{tabular}[c]{@{}l@{}}Exploits advantages of both DQN, ability to learn rich representations,\\ and batch algorithms, stability and data efficiency\end{tabular}                              &                                                                         \\ \cline{1-3}
DQfD {[}10{]}               & Learns from demonstrations                                                                                                                                                                                    & Learns at an accelerated rate from the start.                                                                                                                                                         &                                                                         \\ \cline{1-3}
DQV {[}11{]}                & \begin{tabular}[c]{@{}l@{}}Uses temporal difference to train a Value network\\ and subsequently uses it for training a Quality-Value\\ network that estimates state-action values\end{tabular} & Learns significantly better and faster than DQN and DDQN                                                                                                                                              &                                                                         \\ \cline{1-3}
RUDDER {[}12{]}             & Reward redistribution and return decomposition                                                                                                                                                                & \begin{tabular}[c]{@{}l@{}}Provides prominent improvement on games having long\\  delayed rewards\end{tabular}                                                                                        &                                                                         \\ \cline{1-3}
Ape-X DQfD {[}13{]}         & \begin{tabular}[c]{@{}l@{}}Employs transformed Bellman operator \\ together with temporal consistency loss\end{tabular}                                                                                       & \begin{tabular}[c]{@{}l@{}}Surpasses average human performance on 40 out\\  of 42 Atari 2600   games\end{tabular}                                                                                     &                                                                         \\ \cline{1-3}
Soft DQN {[}14{]}           & \begin{tabular}[c]{@{}l@{}}Incorporation of soft KL penalty and entropy bonus\end{tabular}                                                                                                                 & \begin{tabular}[c]{@{}l@{}}Establishes equivalence between Soft DQN and policy gradient\end{tabular}                                                                                               &                                                                         \\ \hline
DRQN, DARQN {[}14-15{]}           & \begin{tabular}[c]{@{}l@{}}Memory, attention\end{tabular}                                                                                                                 & \begin{tabular}[c]{@{}l@{}}DQN policies modelled by attention-based recurrent networks \end{tabular}                                                                                               &                                                                         \\ \hline\hline
\multicolumn{4}{|c|}{\textbf{Policy Gradient-based DRL}}                                                                                                                                                                                                                                                                                                                                                                                                                                                                  \\ \hline\hline
A3C {[}16{]}                & Asynchronous gradient descent                                                                                                                                                                                 & Consumes less resources; able to run on a standard multi-core CPU                                                                                                                                     & \multirow{3}{*}{on-policy}                                              \\ \cline{1-3}
GA3C {[}17{]}               & Hybrid CPU/GPU-based A3C                                                                                                                                                                                      & Achieves speed significantly higher than its CPU-based counterpart                                                                                                                                    &                                                                         \\ \cline{1-3}
PAAC {[}18{]}               & \begin{tabular}[c]{@{}l@{}}Novel framework for asynchronous algorithms\end{tabular}                                                                                                                        & \begin{tabular}[c]{@{}l@{}}Computationally efficient \& enables faster convergence to optimal policies\end{tabular}                                                                               &                                                                         \\ \hline
PGQL {[}19{]}               & \begin{tabular}[c]{@{}l@{}}Combines on-policy policy gradient\\  with off-policy Q-learning\end{tabular}                                                                                                      & Enhanced stability and data efficiency                                                                                                                                                                & \multirow{3}{*}{off-policy}                                             \\ \cline{1-3}
Retrace($\lambda$) {[}20{]}         & \begin{tabular}[c]{@{}l@{}}Expresses three off-policy algorithms---IS, \\ Q($\lambda$) and TB($\lambda$)--- in a common form\end{tabular}                                                                                     & Safe, sample efficient and has low variance                                                                                                                                                           &                                                                         \\ \cline{1-3}
Reactor {[}21{]}            & Retrace-based actor-critic agent architecture                                                                                                                                                                 & Yields substantial gains in terms of training time.                                                                                                                                                   &                                                                         \\ \hline
IMPALA {[}22{]}             & \begin{tabular}[c]{@{}l@{}}Scalable distributed agent capable\\  of handling multiple tasks with \\ a single set of parameters\end{tabular}                                                                   & \begin{tabular}[c]{@{}l@{}}outperforms state-of-the-art agents in a\\  diverse multi-task environment\end{tabular}                                                                                    & \multirow{3}{*}{on-policy}                                              \\ \cline{1-3}
TRPO {[}23{]}               & \begin{tabular}[c]{@{}l@{}}Employs fixed KL divergence constraint\\  for optimising stochastic control policies\end{tabular}                                                                                  & \begin{tabular}[c]{@{}l@{}}Performs well over a wide variety of large-scale tasks\end{tabular}                                                                                                     &                                                                         \\ \cline{1-3}
PPO {[}24{]}                & \begin{tabular}[c]{@{}l@{}}Makes use of adaptive KL  penalty coefficient\end{tabular}                                                                                                                       & \begin{tabular}[c]{@{}l@{}}As reliable and stable as TRPO but relatively better\\  in terms of implementation and sample complexity\end{tabular}                                                      &                                                                         \\ \hline\hline
\multicolumn{4}{|c|}{\textbf{Model-based DRL}}                                                                                                                                                                                                                                                                                                                                                                                                                                                                            \\ \hline\hline
SimPLe {[}26{]}             & \begin{tabular}[c]{@{}l@{}}Video prediction-based model-based algorithm that\\ requires much fewer  agent-environment\\ interactions than model-free algorithms\end{tabular}                              & \begin{tabular}[c]{@{}l@{}}Outperforms state-of-the-art model-free  algorithms in Atari games\end{tabular}                                                                                          & \multirow{3}{*}{on-policy}                                              \\ \cline{1-3}
TreeQN {[}27{]}             & \begin{tabular}[c]{@{}l@{}}Estimates Q-values based on a dynamic\\  tree constructed   recursively through \\ an implicit transition model\end{tabular}                                                       & \begin{tabular}[c]{@{}l@{}}Outperforms n-step DQN and value prediction networks\\ in multiple Atari games\end{tabular}                                                                               &                                                                         \\ \cline{1-3}
STRAW {[}29{]}              & \begin{tabular}[c]{@{}l@{}}Capable of natural decision making\\  by learning macro actions\end{tabular}                                                                                                       & Improves performance significantly in Atari games                                                                                                                                                     &                                                                         \\ \hline
VProp {[}30{]}              & \begin{tabular}[c]{@{}l@{}}A set of Value Iteration-based planning \\ modules that is trained using RL\end{tabular}                                                                                           & \begin{tabular}[c]{@{}l@{}}\tabitem Able to solve an unseen task and navigate in complex environments\\ \tabitem Able to generalise in dynamic and noisy environment\end{tabular}                            & -                                                                       \\ \hline
MuZero {[}31{]}             & \begin{tabular}[c]{@{}l@{}}Combines tree-based search with learned\\  model to render   superhuman performance\\  in challenging environments\end{tabular}                                                    & \begin{tabular}[c]{@{}l@{}}Delivers state-of-the-art performance on 57 diverse Atari games\end{tabular}                                                                                            & off-policy                                                              \\ \hline
\end{tabular}
\end{table*}

\begin{figure*}[!ht]
\centering
\includegraphics[width=0.85\textwidth]{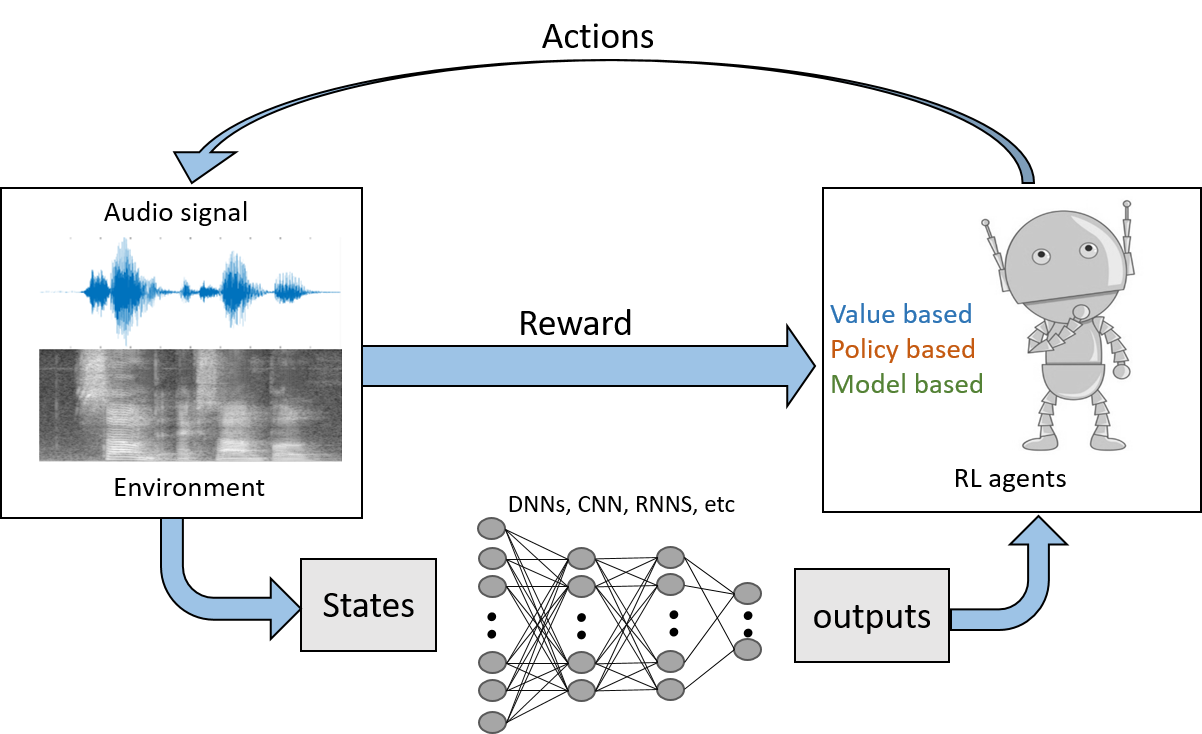}
\caption{Schematic diagram of DRL agents for audio-based applications, where the DL model (via DNNs, CNN, RNNs, etc.) generates audio features from raw waveforms or other audio representations for taking actions that change the environment from state $s_t$ to a next state $s_{t+1}$.}
\label{fig:DRL}
\end{figure*}

\subsection{Model-Based DRL}
Model-based DRL algorithms rely on models of the environment (i.e. underlying dynamics and reward functions) in conjunction with a planning algorithm. Unlike model-free DRL methods that typically entail a large number of samples to render adequate performance, model-based algorithms generally lead to improved sample and time efficiency~\cite{ravindran2019introduction}. 

Kaiser et al.~\cite{kaiser2019model} propose simulated policy learning (SimPLe), a video prediction-based model-based DRL algorithm that requires much fewer agent-environment interactions than model-free algorithms. Experimental results indicate that SimPLe outperforms state-of-the-art model-free algorithms in Atari games. Farquhar et al.~\cite{whiteson2018treeqn} propose TreeQN for complex environments, where the transition model is not explicitly given. The proposed algorithm combines model-free and model-based approaches in order to estimate Q-values based on a dynamic tree constructed recursively through an implicit transition model. Authors of~\cite{whiteson2018treeqn} also propose an actor-critic variant named ATreeC that augments TreeQN with a softmax layer to form a stochastic policy network. They show that both algorithms yield superior performance than n-step DQN and value prediction networks~\cite{oh2017value} on multiple Atari games. Authors in~\cite{vezhnevets2016strategic} introduce a Strategic Attentive Writer (STRAW), which is capable of making natural decisions by learning macro-actions. Unlike state-of-the-art DRL algorithms that yield only one action after every observation, STRAW generates a sequence of actions, thus leading to structured exploration. Experimental results indicate a significant improvement in Atari games with STRAW. Value Propagation (VProp)~\cite{nardelli2018value} is a set of Value Iteration-based planning modules trained using RL and capable of solving unseen tasks and navigating in complex environments. It is also demonstrated that VProp is able to generalise in a dynamic and noisy environment. Authors in~\cite{schrittwieser2019mastering} present a model-based algorithm named MuZero that combines tree-based search with a learned model to render superhuman performance in challenging environments. Experimental results demonstrate that MuZero delivers state-of-the-art performance on 57 diverse Atari games. Table ~\ref{Summary of DRL algorithms} presents an overview of DRL algorithms for a reader's glance.

\section{Audio-Based DRL}
\label{survey}
This section surveys related works where audio is a key element in the learning environments of DRL agents. An example scenario is a human speaking to a machine trained via DRL as in Figure~\ref{fig:DRL}, where the machine has to act based on features derived from audio signals. Table~\ref{DRL:summ} summarises the characterisation of DRL agents for six audio-related areas: 
\begin{enumerate}
\item automatic speech recognition; 
\item spoken dialogue systems; 
\item emotions modelling; 
\item audio enhancement; 
\item music listening and generation; and 
\item robotics, control and interaction.
\end{enumerate}

\begin{table*}[!ht]
\centering
\caption{Summary of audio related fields, characterisation of DRL agents, and related datasets.}
\begin{tabular}{|l|l|l|}
\hline
{\bf Appl. Area}                                                                          & {\bf State Representations $\mathcal{S}$, Actions $\mathcal{A}$, and Reward Functions $\mathcal{R}$}                              & {\bf Popular Datasets}                              \\ \hline\hline
\begin{tabular}[c]{@{}l@{}}Automatic \\Speech \\ Recognition\end{tabular} & \begin{tabular}[c]{@{}l@{}}$\mathcal{S}$: States are learnt representations from input speech features (e.g. fMLLR or MFCC vectors~\cite{RathPVC13})\vspace{1mm}. \\$\mathcal{A}$: Actions include phonemes, graphemes, commands, or candidates from the ASR N-best list.\vspace{1mm}\\$\mathcal{R}$: They have included binary rewards (positive for selecting the correct choice, 0 otherwise), \\and non-binary sentence/token rewards based on the Levenshtein distance algorithm.\vspace{1mm}\end{tabular}    & \begin{tabular}[c]{@{}l@{}}\tabitem LibriSpeech~\cite{panayotov2015librispeech}\\ \tabitem TED-LIUM~\cite{rousseau2012ted} \\ \tabitem  Wall Street Journal~\cite{paul1992design}\\ \tabitem SWITCHBOARD~\cite{godfrey1992switchboard}\\ \tabitem TIMIT~\cite{garofolo1993darpa}\end{tabular} \\ \hline
\begin{tabular}[c]{@{}l@{}}Spoken \\Dialogue \\ Systems\end{tabular} & \begin{tabular}[c]{@{}l@{}}$\mathcal{S}$: They encode the uttered words by the system and recognised user words into a dialogue history \\and some additional information from classifiers such as user goals, user intents, speech recognition \\confidence scores, and visual information (in the case of multimodal systems), among others. \vspace{1mm}\\$\mathcal{A}$: While actions in task-oriented systems include slot requests/confirmations/apologies, slot-value\\ selection, ask question, data retrieval, information presentation, among others; actions in open-ended \\systems include either all possible sentences (infinite) or clusters of sentences (finite). \vspace{1mm}\\$\mathcal{R}$: They vary depending on the company/project requirements and tend to include sparse and \\non-sparse numerical rewards such as dialogue length, task success, dialogue similarity, dialogue \\coherence, dialogue repetitiveness, game scores (in the case of game-based systems), among others.\vspace{1mm}\end{tabular} & \begin{tabular}[c]{@{}l@{}} \tabitem  SGD~\cite{RastogiZSGK20}\\\tabitem DSTC~\cite{WilliamsRH16a}\\\tabitem Frames~\cite{AsriSSZHFMS17}\\\tabitem MultiWOZ~\cite{budzianowski2018multiwoz}\\\tabitem SubTle Corpus~\cite{ameixa2013subtitles}\\\tabitem Simulations \cite{SchatzmannWSY06}\\\tabitem Other datasets~\cite{SerbanLHCP18}\end{tabular} \\ \hline
\begin{tabular}[c]{@{}l@{}}Speech \\Emotion \\ Recognition\end{tabular}& \begin{tabular}[c]{@{}l@{}}
$\mathcal{S}$: Speech features (e.g., MFCC) are considered as input features.\vspace{1mm}\\
$\mathcal{A}$: Actions include speech emotion labels (e.g. unhappy, neutral, happy), sentiment detection \\(e.g. negative, neutral, positive), and termination from utterance listening. \vspace{1mm}\\
$\mathcal{R}$: Binary reward functions have been used (positive for choosing the correct choice, 0 otherwise). \vspace{1mm}\end{tabular}    & \begin{tabular}[c]{@{}l@{}}\tabitem EMODB~\cite{burkhardt2005database}\\ \tabitem IEMOCAP~\cite{busso2008iemocap}\\ \tabitem MSP-IMPROV~\cite{busso2016msp}\\ \tabitem SEMAINE~\cite{mckeown2011semaine} \\ \tabitem MELD~\cite{poria2019meld}\end{tabular} \\ \hline
\begin{tabular}[c]{@{}l@{}}Audio \\Enhancement\end{tabular}& \begin{tabular}[c]{@{}l@{}}
$\mathcal{S}$: States are learnt from clean and noisy acoustic features. \vspace{1mm}\\
$\mathcal{A}$: Finding closest cluster and its index, time-frequency mask estimation, and increasing or decreasing\\ the parameter values of the speech-enhancement algorithm.\vspace{1mm}\\
$\mathcal{R}$: Positive rewards for correct choice, negative otherwise.\vspace{1mm}\end{tabular}    & \begin{tabular}[c]{@{}l@{}} \tabitem DEMAND~\cite{thiemann2013diverse}\\ \tabitem CHiME-3~\cite{barker2015third}\\ \tabitem WHAMR~\cite{maciejewski2020whamr}\end{tabular} \\ \hline
\begin{tabular}[c]{@{}l@{}}Music \\Generation\end{tabular} & \begin{tabular}[c]{@{}l@{}}
$\mathcal{S}$: State representations are learned from Musical notes.\vspace{1mm}\\ 
$\mathcal{A}$: Musical generation and next note selection are considered as actions. \vspace{1mm}\\
$\mathcal{R}$: Binary reward functions based on hard-coded musical theory rules, including the likelihood of actions. \vspace{1mm}

\end{tabular}    & \begin{tabular}[c]{@{}l@{}} \tabitem  Classical piano MIDI\\ database~\cite{krueger2016classical}\\ \tabitem MusicNet dataset~\cite{thickstun2016learning}\\ \tabitem JSB Chorales \\dataset~\cite{allan2005harmonising}\end{tabular} \\ \hline
\begin{tabular}[c]{@{}l@{}}Robotics, \\Control and \\Interaction\end{tabular}& \begin{tabular}[c]{@{}l@{}}$\mathcal{S}$: They encode visual and verbal  representations derived from image embeddings, speech features, \\and word or sentence embeddings. Additional information include user intents, speech recognition \\scores, human activities, postures, emotions, and body joint angles, among others. \vspace{1mm}\\$\mathcal{A}$: They include motor commands (e.g. gestures, locomotion, navigation, manipulation, gaze) and \\verbalizations such as dialogue acts and backchannels (e.g. laughs, smiles, noddings, head-shakes).\vspace{1mm}\\$\mathcal{R}$: They are based on task success (positive rewards for achieving the goal, negative rewards for \\failing the task, and zero/shaped rewards otherwise) and user engagement. \end{tabular}    & \begin{tabular}[c]{@{}l@{}} \tabitem AVDIAR~\cite{gebru2017audio}\\ \tabitem NLI Corpus~\cite{ScaliseLARS18}\\ \tabitem VEIL dataset~\cite{MisraSLS16} \\\tabitem Simulations~\cite{Cuayahuitl20}\\ \tabitem Real-world \\interactions~\cite{QureshiNYI18}\end{tabular} \\ \hline
\end{tabular}
\label{DRL:summ}
\end{table*}

 \subsection{Automatic Speech Recognition (ASR)}
Automatic speech recognition (ASR) is the process of converting a speech signal into its corresponding text by algorithms. Contemporary ASR technology has reached great levels of performance due to advancements in DL models. The performance of ASR systems, however, relies heavily on supervised training of deep models with large amounts of transcribed data. Even for resource-rich languages, additional transcription costs required for new tasks hinders the applications of ASR. To broaden the scope of ASR, different studies have attempted RL based models with the ability to learn from feedback. This form of learning aims to reduce transcription costs and time by humans providing positive or negative rewards instead of detailed transcriptions. For instance, Kala et al.~\cite{kala2018reinforcement} proposed an RL framework for ASR based on the policy gradient method that provides a new view of existing training and adaptation methods. They achieved improved recognition performance and reduced Word Error Rate (WER) compared to unsupervised adaptation. In ASR, sequence-to-sequence models have shown great success; however, these models fail to approximate real-world speech during inference. Tjandra et al.~\cite{tjandra2018sequence} solved this issue by training a sequence-to-sequence model with a policy gradient algorithm. Their results showed a significant improvement using an RL-based objective and a maximum likelihood estimation (MLE) objective compared to the model trained with only the MLE objective. In another study,~\cite{tjandra2019end} extended their own work by providing more details on their model and experimentation. They found that using token-level rewards (intermediate rewards are given after each time step) provide improved performance compared to sentence-level rewards and baseline systems. In order solve the issues of semi-supervised training of sequence-to-sequence ASR models, Chung et al. \cite{chung2020semi} investigated the REINFORCE algorithm by rewarding the
ASR to output more correct sentences for both unpaired and paired speech input data. Experimental evaluations showed that the DRL-based method was able to effectively reduce character error rates from 10.4\% to 8.7\%.

Karita et al.~\cite{karita2018sequence} propose to train an encoder-decoder ASR system using a sequence-level evaluation metric based on the policy gradient objective function. This enables the minimisation of the expected WER of the model predictions. In this way, the authors found that the proposed method improves recognition performance.
%In~\cite{shannon2017optimizing}, the author used the policy gradient framework to optimize WER for the acoustic model using connectionist temporal classification (CTC). Results showed that the proposed model achieves  5\% relative improvement in WER over a well-tuned state-level minimum Bayes risk (sMBR) baseline. 
The ASR system of~\cite{zhou2018improving} was jointly trained with maximum likelihood and policy gradient to improve via end-to-end learning. They were able to optimise the performance metric directly and achieve 4\% to 13\% relative performance improvement. In \cite{luo2017learning}, the authors attempted to solve sequence-to-sequence problems by proposing a model based on supervised backpropagation and a policy gradient method, which can directly maximise the log probability of the correct answer. They achieved very encouraging results on a small scale and a medium scale ASR. Radzikowski et al.~\cite{radzikowski2019dual} proposed a dual supervised model based on a policy gradient methodology for non-native speech recognition. They were able to achieve promising results for the English language pronounced by Japanese and Polish speakers. % based on the results.

To achieve the best possible accuracy, end-to-end ASR systems are becoming increasingly large and complex. DRL  methods can also be leveraged to provide model compression~\cite{he2018amc}. In~\cite{dudziak2019shrinkml}, RL-based ShrinkML is proposed to optimise the per-layer compression ratios in a state-of-the-art LSTM-based ASR model with attention. For time-efficient ASR,~\cite{rajapakshe2020deep} evaluated the pre-training of an RL-based policy gradient network. They found that pre-training in DRL offers faster convergence compared to non-pre-trained networks, and also achieve improved recognition in lesser time. To tackle the slow convergence time of the REINFORCE algorithm~\cite{williams1992simple}, Lawson et al.~\cite{lawson2018learning}, evaluated Variational Inference for Monte Carlo Objectives (VIMCO) and Neural Variational Inference (NVIL) for phoneme recognition tasks in clean and noisy environments. The authors found that the proposed method (using VIMCO and NVIL) outperforms REINFORCE and other methods at training online sequence-to-sequence models. 

All of the above-mentioned studies highlight several benefits of using DRL for ASR. Despite these promising results, further research is required on DRL algorithms towards building autonomous ASR systems that can work in complex real-life settings. REINFORCE algorithm is very popular in ASR, therefore, research is also required to explore other DRL algorithms to highlights suitability for ASR.

\subsection{Spoken Dialogue Systems (SDSs)}
Spoken dialogue systems are gaining interest due to many applications in customer services and goal-oriented human-computer-interaction. Typical SDSs integrate several key components including speech recogniser, intent recogniser, knowledge base and/or database backend, dialogue manager, language generator, and speech synthesis, among others~\cite{ZueGlass2000}. The task of a dialogue manager in SDSs is to select actions based on observed events~\cite{LevinPE00,singh2000reinforcement}. Researchers have shown that the action selection process can be effectively optimised using RL to model the dynamics of spoken dialogue as a fully or partially observable Markov Decision Process~\cite{paek2006reinforcement}. Numerous studies have utilised RL-based algorithms in spoken dialogue systems. In contrast to text-based dialogue systems that can be trained directly using large amounts of text data \cite {GaoGL19}, most SDSs have been trained using user simulations~\cite{SchatzmannWSY06}. The justification for that is mainly due to insufficient amounts of training dialogues to train or test from real data~\cite{SerbanLHCP18}.
% (which have been trained from small dialogue data)

%But given the amount of training data required by RL-based methods, many of these systems have been trained using user simulations~\cite{SchatzmannWSY06} due to insufficient training dialogues~\cite{SerbanLHCP18}. 

SDSs involve policy optimisation to respond to humans by taking the current state of the dialogue, selecting an action,  and returning the verbal response of the system. For instance, Chen et al. \cite{chen2020deep} presented an online DRL-based dialogue state tracking framework in order to improve the performance of a dialogue manager. They achieved promising results for online dialogue state tracking in the second and third dialog state
tracking challenges ( \cite{henderson2014second,henderson2014third}). Weisz et al.~\cite{weisz2018sample} utilised DRL approaches, including actor-critic methods and off-policy RL. They also evaluated actor-critic with experience replay (ACER)~\cite{wang2016sample,munos2016safe}, which has shown promising results on simple gaming
tasks. They showed that the proposed method is sample efficient and that performed better than some state-of-the-art DL approaches for spoken dialogue. A task-oriented end-to-end DRL-based dialogue system is proposed in~\cite{cuayahuitl2017simpleds}. They showed that DRL-based
optimisation produced significant improvement in task success rate and also caused a reduction in dialogue length compared to supervised training. Zhao et al.~\cite{zhao2016towards} utilised deep recurrent Q-networks (DRQN) for dialogue state tracking and management. Experimental results showed that the proposed model can exploit the strengths of DRL and supervised learning to achieve faster learning speed and better results than the modular-based baseline system. To present baseline results, a benchmark study~\cite{casanueva2017benchmarking} is performed using DRL algorithms including DQN, A2C and natural actor-critic~\cite{su2017sample} and their performance is compared against GP-SARSA~\cite{gavsic2013gaussian}. Based on experimental results on the PyDial toolkit~\cite{ultes2017pydial}, the authors conclude that substantial improvements are still needed for DRL methods to match the performance of carefully designed handcrafted policies.  
In addition to SDSs optimised via flat DRL, hierarchical RL/DRL methods have been proposed for policy learning using dialogue states with different levels of abstraction and dialogue actions at different levels of granularity (via primitive and composite actions)~\cite{cuayahuitl_thesis2009,CuayahuitlRLS10,DethlefsC15,BudzianowskiUSM17,PengLLGCLW17,ZhangZY18}. The benefits of this form of learning include faster training and policy reuse. A deep Q-network based multi-domain dialogue system is proposed in~\cite{cuayahuitl2016deep}. They train the proposed SDS using a network of DQN agents, which is similar to hierarchical DRL but with more flexibility for transitioning across dialogues domains. Another work related to faster training is proposed by~\cite{Gordon-HallGC20}, where the behaviour of RL agents is guided by expert demonstrations. 

The optimisation of dialogue policies requires a reward function that unfortunately is not easy to specify. %an accurate reward function. 
This often requires annotated data for training a reward predictor instead of a hand-crafted one. In real-world applications, such annotations are either scarce or not available. Therefore, some researchers have turner their attention to methods for online active reward learning. In~\cite{su2016line}, the authors presented an online learning framework for a spoken dialogue system. They jointly trained the dialogue policy alongside the reward model via active learning. Based on the results, the authors showed that the proposed framework can significantly reduce data annotation costs and can also mitigate noisy user feedback in dialogue policy learning. Su et al.~\cite{su2017sample} introduced two approaches: trust region actor-critic with experience replay (TRACER) and episodic natural actor-critic with experience replay (eNACER) for dialogue policy optimisation. From these two algorithms, they achieved the best performance using TRACER.

In~\cite{ultes2017domain}, the authors propose to learn a domain-independent reward function based on user satisfaction for dialogue policy learning. The authors showed that the proposed framework yields good performance for both task success rate and user satisfaction. Researchers have also used DRL to learn dialogue policies in noisy environments, and some have shown that their proposed models can generate dialogues indistinguishable from human ones~\cite{fazel2017learning}. Carrara et al.~\cite{carrara2017online} propose online learning and transfer for user adaptation in RL-based dialogue systems. Experiments were carried out on a negotiation dialogue task, which showed significant improvements over baselines. In another study~\cite{carrara2018safe}, authors proposed $\epsilon$-safe, a Q-learning algorithm, for safe transfer learning for dialogue applications. A DRL-based chatbot called MILABOT was designed in~\cite{serban2017deep}, which can converse with humans on popular topics through both speech and text---performing significantly better than many competing systems. The text-based chatbot  in~\cite{CuayahuitlLRCCI19} used an ensemble of DRL agents, and showed that training multiple dialogue agents performs better than a single agent. 

\begin{table*}[!ht]
\scriptsize
\caption{Summary of research papers on dialogue systems trained with DRL algorithms (?=information not available)}
\centering
\begin{tabular}{|c|l|c|c|c|c|c|l|}
%{|l|l|l|l|l|l|l|l|}
\hline
{\bf Refe-} & {\bf Application} & {\bf DRL } & {\bf User Si-}            & {\bf Transfer}  & {\bf Training}      & {\bf Human}      & {\bf Reward} \\
{\bf rence} & {\bf Domain(s)}  &  {\bf Algorithm}                       & {\bf mulations} & {\bf Learning} & {\bf (Test) Data} & {\bf Evaluation} & {\bf Function} \\
\hline
\hline
\multirow{2}{*}{\cite{AmmanabroluR19}} & Games:  & \multirow{2}{*}{KG-DQN} & \multirow{2}{*}{No} & \multirow{2}{*}{Yes} & \multirow{2}{*}{40 (10) games} & \multirow{2}{*}{No} & +1 for getting closer to the finish, -1 for\\
& Slice of Life, Horror &  & & & & & extending the minimum steps, 0 otherwise\\
\hline
\multirow{2}{*}{\cite{CasanuevaBSURTG18}} & \multirow{2}{*}{Restaurants, laptops} & FDQN,  & \multirow{2}{*}{Yes} & \multirow{2}{*}{No} & \multirow{2}{*}{4K ($0.5$K) dialogues} & \multirow{2}{*}{No} & +20 if successful dialogue or 0 otherwise,\\ 
& & GP-Sarsa & & & & & minus dialogue length\\
\hline
\multirow{2}{*}{\cite{ChenCCTGY18}} & \multirow{2}{*}{Restaurants} & \multirow{2}{*}{MADQN} & \multirow{2}{*}{Yes} & \multirow{2}{*}{Yes} & \multirow{2}{*}{15K ($?$) dialogues} & \multirow{2}{*}{No} & +1 if successful dialogue, \\
& &  & & & & & -0.05 at each dialogue turn\\
\hline
\multirow{2}{*}{\cite{Cuayahuitl20}} & \multirow{2}{*}{Chitchat} & Ensemble & \multirow{2}{*}{No} & \multirow{2}{*}{No} & \multirow{2}{*}{$\leq$64K (1k) dialogues} & \multirow{2}{*}{Yes} & +1 for a human-like response, \\
& & DQN & & & & & -1 for a randomly chosen response\\
\hline
\multirow{2}{*}{\cite{CuayahuitlLRCCI19}} & Robot playing & Competitive & \multirow{2}{*}{Yes} & \multirow{2}{*}{No} & \multirow{2}{*}{20K (3K) games}  & \multirow{2}{*}{Yes} & +5 for a game win, \\
& noughts \& crosses & DQN & & & & & +1 for a draw, -5 for a loss\\
\hline
\multirow{2}{*}{\cite{CuayahuitlYWC17}} & \multirow{2}{*}{Restaurants, hotels} & \multirow{2}{*}{NDQN} & \multirow{2}{*}{Yes} & \multirow{2}{*}{No} & \multirow{2}{*}{8.7K (1K) dialogues} & \multirow{2}{*}{No} & $Pr$(TaskSuccess) plus $Pr$(Data-Like)\\
& & & & & & & minus number of turns $\times -0.1$\\
\hline
\multirow{2}{*}{\cite{DasKMLB17}} & Visual & \multirow{2}{*}{Reinforce} & \multirow{2}{*}{No} & \multirow{2}{*}{No} & \multirow{2}{*}{68K (9.5K) images} & \multirow{2}{*}{Yes} & Euclidean distances between predicted and \\
& Question-Answering & & & & & & target descriptions of the last 2 time steps\\
\hline
\multirow{2}{*}{\cite{FatemiASHS16}} & \multirow{2}{*}{Restaurants} & \multirow{2}{*}{DA2C} & \multirow{2}{*}{Yes} & \multirow{2}{*}{No} & \multirow{2}{*}{15K (0.5K) dialogues} &  \multirow{2}{*}{No} & +1 if successful dialogue, -0.03 at each turn, \\
& & & & & & & -1 if unsuccessful dialogue or hangup\\
\hline
\multirow{2}{*}{\cite{abs-1712-04034}} & \multirow{2}{*}{Movie chat} & Dueling & \multirow{2}{*}{Yes} & \multirow{2}{*}{No} & \multirow{2}{*}{150K ($?$) sentences} & \multirow{2}{*}{No} & +10 for correct recognition, -12 for incorrect\\
& & DDQN & & & & & recognition, smaller rewards for confirm/elicit\\
\hline
\multirow{2}{*}{\cite{Gordon-HallGC20}} & \multirow{2}{*}{MultiWoz} & \multirow{2}{*}{NDfQ} & \multirow{2}{*}{Yes} & \multirow{2}{*}{No} & \multirow{2}{*}{11.4K (1K) dialogues} & \multirow{2}{*}{No} & +100 for successfully completing the task, \\
& & & & & & & -1 at each turn\\
\hline
\multirow{2}{*}{\cite{abs-1907-00456}} & \multirow{2}{*}{Chitchat} & \multirow{2}{*}{DBCQ} & \multirow{2}{*}{No} & \multirow{2}{*}{No} & \multirow{2}{*}{14.2$\times$2K ($?$) sentences} & {Yes} & Weighted scores combining sentiment, asking,\\
& & & & & & & laughter, long dialogues, \& sentence similarity\\
\hline
\multirow{2}{*}{\cite{LiMRGGJ16}} & \multirow{2}{*}{OpenSubtitles} & \multirow{2}{*}{Reinforce} & \multirow{2}{*}{No} & \multirow{2}{*}{No} & \multirow{2}{*}{\~10M (1K) sentences} & \multirow{2}{*}{Yes} & Weighted scores combining ease of answering,\\
& & & & & & & information flow, and semantic coherence\\
\hline
\multirow{2}{*}{\cite{LiMSJRJ17}} & \multirow{2}{*}{OpenSubtitles}  & Adversarial & \multirow{2}{*}{No} & \multirow{2}{*}{No} & \multirow{2}{*}{$?$ ($?$) dialogues} & \multirow{2}{*}{Yes} & Learnt rewards (binary classifier determining\\
& & Reinforce & & & & & a machine- or human-generated dialogue)\\
\hline
\multirow{2}{*}{\cite{LiptonLG00D18}} & \multirow{2}{*}{Movie booking} & \multirow{2}{*}{BBQN} & \multirow{2}{*}{Yes} & \multirow{2}{*}{No} & \multirow{2}{*}{20K (10K) dialogues} & \multirow{2}{*}{Yes} & +40 if successful dialogue, -1 at each turn, \\
& & & & & & & -10 for a failed dialogue\\
\hline
\multirow{2}{*}{\cite{NarasimhanBJ18}} & Freeway, Bomberman, & Text-DQN, & \multirow{2}{*}{No} & \multirow{2}{*}{Yes} & \multirow{2}{*}{10M-15M (50K) steps} & \multirow{2}{*}{No} & Learnt rewards (CNN network trained from\\
       & Bourderchase, F\&E  & Text-VI &  &  &  &  & crowdsourced text descriptions of gameplays)\\
\hline
\multirow{2}{*}{\cite{PengLLGCLW17}} & \multirow{2}{*}{Flights and hotels} & Hierarchical  & \multirow{2}{*}{Yes} & \multirow{2}{*}{No} & \multirow{2}{*}{20K (2K) dialogues} & \multirow{2}{*}{Yes} & +120 if successful dialogue, -1 at each turn,\\
& & DQN & & & & & -60 for a failed dialogue\\
\hline
\multirow{2}{*}{\cite{PengLGLCW18}} & \multirow{2}{*}{Movie-ticket booking} & Adversarial & \multirow{2}{*}{Yes} & \multirow{2}{*}{No} & \multirow{2}{*}{100K (5K) dialogues} & \multirow{2}{*}{No} & Learnt rewards (MLP network comparing\\
& & A2C & & & & & state-action pairs with human dialogues)\\
\hline
\multirow{2}{*}{\cite{SalehJGSP20}} & \multirow{2}{*}{Chitchat} & Hierarchical & \multirow{2}{*}{No} & \multirow{2}{*}{No} & \multirow{2}{*}{109K (10K) dialogues} & \multirow{2}{*}{Yes} & Predefined scores combining question, \\
& & Reinforce & & & & & repetition, semantic similarity, and toxicity\\
\hline
\multirow{2}{*}{\cite{SankarR19}} & \multirow{2}{*}{Chitchat} & \multirow{2}{*}{Reinforce} & \multirow{2}{*}{No} & \multirow{2}{*}{No} & \multirow{2}{*}{$\sim$2M ($?$) dialogues} & \multirow{2}{*}{Yes} & Positive reward from ease of answering - \\
& & & & & & & negative reward for manual dull utterances\\
\hline
\multirow{2}{*}{\cite{abs-1709-02349}} & \multirow{2}{*}{Chitchat} & \multirow{2}{*}{Reinforce} & \multirow{2}{*}{No} & \multirow{2}{*}{No} & \multirow{2}{*}{$\sim$5K (0.1) dialogues} & \multirow{2}{*}{Yes} & Learnt rewards (linear regressor predicting\\
& & & & & & & user scores at the end of the dialogue)\\
\hline
\multirow{2}{*}{\cite{SuBUGY17}} & \multirow{2}{*}{Restaurants} & TRACER,  & \multirow{2}{*}{Yes} & \multirow{2}{*}{No} & \multirow{2}{*}{$\leq$3.5K (0.6K) dialogues} & \multirow{2}{*}{No} & +20 if successful dialogue (0 otherwise)\\ 
& & eNACER & & & & & minus 0.05 $\times$ number of dialogue turns\\
\hline
\multirow{2}{*}{\cite{UltesBCMRSWGY17}} & Buses, restaurants, & \multirow{2}{*}{GP-Sarsa} & \multirow{2}{*}{Yes} & \multirow{2}{*}{No} & \multirow{2}{*}{1K (0.1K) dialogues} & \multirow{2}{*}{No} & Learnt rewards (Support Vector Machine\\
& hotels, laptops & & & & & & predicting user dialogue ratings)\\
\hline
\multirow{2}{*}{\cite{XuZGLTL19}} & Medical diagnosis & \multirow{2}{*}{KR-DQN} & \multirow{2}{*}{Yes} & \multirow{2}{*}{No} & \multirow{2}{*}{423 (104) dialogues} & \multirow{2}{*}{Yes} & +44 for successful diagnoses, -22 for failed\\
& (4 diseases) & & & & & & diagnoses, -1 for failing symptom requests\\
\hline
\multirow{2}{*}{\cite{abs-1802-03753}} & \multirow{2}{*}{Restaurants} & \multirow{2}{*}{ACER} & \multirow{2}{*}{Yes} & \multirow{2}{*}{No} & \multirow{2}{*}{4K (4K) dialogues} & \multirow{2}{*}{Yes} & +20 for a successful dialogue minus number\\
& & & & & & & of turns in the dialogue\\
\hline
\multirow{2}{*}{\cite{WilliamsZ16}} & \multirow{2}{*}{Dialling domain} & \multirow{2}{*}{Reinforce} & \multirow{2}{*}{Yes} & \multirow{2}{*}{No} & \multirow{2}{*}{5K (0.5K) dialogues} & \multirow{2}{*}{No} & +1 for successfully completing the dialogue, \\
& & & & & & & 0 otherwise\\
\hline
\multirow{2}{*}{\cite{ZhaoE16}} & \multirow{2}{*}{20-question game} & \multirow{2}{*}{DRQN} & \multirow{2}{*}{Yes} & \multirow{2}{*}{No} & \multirow{2}{*}{120K (5K) sentences} & \multirow{2}{*}{No} & +30 for a game win, -30 for a lost game, \\
& & & & & & & -5 for a wrong guess\\
\hline
\multirow{2}{*}{\cite{ZhaoXE19}} & DealOrNotDeal, & \multirow{2}{*}{Reinforce} & \multirow{2}{*}{Yes;No} & \multirow{2}{*}{No} & \multirow{2}{*}{$\leq$8.4K ($\leq$1K) dialogues} & \multirow{2}{*}{No} & +$\le$10 for a negotiation, 0 for no agreement;\\
& MultiWoz & & & & & & language constrained reward curve\\
\hline
\multirow{2}{*}{\cite{ZhangZY18}} & 20 images  & DRRN+ & \multirow{2}{*}{Yes} & \multirow{2}{*}{No} & \multirow{2}{*}{20K (1K) games} & \multirow{2}{*}{No} & +10 for a game win, -10 for a lost game, \\
& guessing game & DQN & & & & & a pseudo reward for question selection\\
\hline
\multirow{2}{*}{\cite{TakanobuZH19}} & \multirow{2}{*}{MultiWoz}  & GP$_{\mbox{mbcm}}$, PPO & \multirow{2}{*}{Yes} & \multirow{2}{*}{No} & \multirow{2}{*}{10.5K (1K) dialogues} & \multirow{2}{*}{Yes} & Learnt rewards (MLP network comparing\\
& & ACER, ALDM & & & & & state-action pairs with human dialogues)\\
\hline
\end{tabular}
\label{DRL-ALGS}
\end{table*}

Table~\ref{DRL-ALGS} shows a summary of DRL-based dialogue systems. While not all involve spoken interactions, they can be applied to speech-based systems by for example using the outputs from a speech recogniser instead of typed interactions. In terms of application, we can observe that most systems focus on one or a few domains---systems trained with a large amount of domains is usually not attempted, presumably due to the high requirements of data and compute involved. Regarding algorithms, the most popular are DQN-based or REINFORCE among other more recent algorithms---when to use one over another algorithm still needs to be understood better. We can also observe that user simulations are mostly used for training task-oriented dialogue systems, while real data is the preferred choice for open-ended dialogue systems. We can note that while transfer learning is an important component in a trained SDS, it is not common-place yet. Given that learning from scratch every time a system is trained is neither scalable nor practical, it looks like transfer learning will naturally be adopted more and more in the future as more domains are taken into account. In terms of datasets, most of them they are still in the small size. It is rare to see SDSs trained with millions of training dialogues or sentences. As datasets grow, the need for more efficient training methods will take more relevance in future systems. Regarding human evaluations, we can observe that about half of research works involve human evaluations. While human evaluations may not  always be required to answer a research question, they certainly should be used whenever learnt conversational skills are being assessed or judged. We can also note that there is no standard for specifying reward functions due to the wide variety of  functions used in previous works---almost every paper uses a different reward function. Even when some works use learnt reward functions (e.g. based on adversarial learning), they focus on learning to discriminate between machine-generated and human generated dialogues without taking other dimensions into account such as task success or additional penalties. Although there is advancement in the specification of reward functions by learning them instead of hand-crafting them, this area requires better understanding for optimising different types of dialogues including information-seeking, chitchat, game-based, negotiation-based, etc.

\subsection{Emotions Modelling}
Emotions are essential in vocal human communication, and they have recently received growing interest by the research community~\cite{latif2019unsupervised,wanrev,latifdeep}. Arguably, human-robot interaction can be significantly enhanced if dialogue agents can perceive the emotional state of a user and its dynamics~\cite{maasur,majumder2019dialoguernn}. This line of research is categorised into two areas: emotion recognition in conversations~\cite{poria2019emotion}, and affective dialogue generation~\cite{young2020dialogue,zhou2018emotional}. Speech emotion recognition (SER) can be used as a reward for RL based dialogue systems \cite{heusser2019bimodal}. This would allow the system to adjust the behaviour based on the emotional states of the dialogue partner. Lack of labelled emotional corpora and low accuracy in SER are two major challenges in the field. To achieve the best possible accuracy, various DL-based methods have been applied to SER, however, performance improvement is still needed for real-time deployments. DRL offers different advantages to SER, as highlighted in different studies. In order to improve audio-visual SER performance, Ouyang et al. \cite{ouyang2018audio} presented a model-based RL framework that utilised feedback of testing results as rewards from environment to update the fusion weights. They evaluated the proposed model on the Multimodal Emotion Recognition
Challenge (MEC 2017) dataset and achieved top 2 at the MEC 2017 Audio-Visual Challenge. To minimise the latency in SER, Lakomin et al.~\cite{lakomkin2018emorl} proposed EmoRL for predicting the emotional state of a speaker as soon as it gains enough confidence while listening. In this way, EmoRL was able to achieve lower latency and minimise the need for audio segmentation required in DL-based approaches for SER.   In~\cite{sangeetha2019emotion}, authors used RL with an adaptive fractional deep Belief network (AFDBN) for SER to enhance human-computer interaction. They showed that the combination of RL with AFDBN is efficient in terms of processing time and SER performance. Another study~\cite{chen2017multimodal} utilised an LSTM-based gated multimodal embedding with temporal attention for sentiment analysis. They exploited the policy gradient method REINFORCE to balance exploration and optimisation by random sampling. They empirically show that the proposed model was able to deal with various challenges of understanding communication dynamics. 

DRL is less popular in SER compared ASR and SDSs. The above mentioned studies attempted to helps solving different SER challenges using DRL, however, there is still a need for developing adaptive SER agents that can perform SER in cross-lingual settings.

\subsection{Audio Enhancement}
The performance of audio-based intelligent systems is critically vulnerable to noisy conditions and degrades according to the noise levels in the environment~\cite{li2015robust}. Several approaches have been proposed~\cite{li2013investigation} to address problems caused by environmental noise. One popular approach is audio enhancement, which aims to generate an enhanced audio signal from its noisy or corrupted version~\cite{wang2016joint}. DL-based speech enhancement has attained increased attention due to its superior performance compared to traditional methods~\cite{baby2015exemplar,wang2018supervised}.

In DL-based systems, the audio enhancement module is generally optimised separately from the main task such as minimisation of WER. Besides the speech enhancement module, there are different other units in speech-based systems which increase their complexity and make them non-differentiable. In such situations, DRL can achieve complex goals in an iterative manner, which makes it suitable for such applications. Such DRL-based approaches have been proposed in~\cite{shen2019reinforcement} to optimise the speech enhancement module based on the speech recognition results. Experimental results have shown that DRL-based methods can effectively improve the system's performance by 12.4\% and 19.2\% error rate reductions for the signal to noise ratio at 0 dB and 5 dB, respectively. In~\cite{koizumi2017dnn}, authors attempted to optimise DNN-based source enhancement using RL with numerical rewards calculated from conventional perceptual scores such as perceptual evaluation of speech quality (PESQ)~\cite{recommendation2001perceptual} and perceptual evaluation methods for audio source separation (PEASS)~\cite{emiya2011subjective}. They showed empirically that the proposed method can improve the quality of the output speech signals by using RL-based optimisation. Fakoor et al.~\cite{fakoor2017reinforcement} performed a study in an attempt to improve the adaptivity of speech enhancement methods via RL. They propose to model the noise-suppression module as a black box, requiring no knowledge of the algorithmic mechanics. Using an LSTM-based agent, they showed that their method improves system performance compared to methods with no adaptivity. In \cite{9247199}, the authors presented a DRL-based method to achieve personalised compression from noisy speech for a specific user in a hearing aid application. To deal with non-linearities of human hearing via the reward/punishment mechanism, they used a DRL agent that receives preference feedback from the target user. Experimental results showed that the developed approach achieved preferred hearing outcomes. 

Similar to SER, very few studies explored DRL for audio enhancement. Most of these studies evaluated DRL-based methods to achieve a certain level of signal enhancement in a controlled environment. Further research efforts are needed to develop DRL agents that can perform their tasks in real and complex noisy environments.

\subsection{Music Listening and Generation}
DL models are widely used for generating content including images, text, and music. The motivation for using DL for music generation lies in its generality since it can learn from arbitrary corpora of music and be able to generate various musical genres compared to classical methods~\cite{steedman1984generative,ebciouglu1988expert}. 

Here, DRL offers opportunities to impose rules of music theory for the generation of more real musical structures~\cite{jaques2016generating}. Various researchers have explored such opportunities of DRL for music generation. For instance, in~\cite{kotecha2018bach2bach} authors achieved better quantitative and qualitative results using an LSTM-based architecture in an RL setting generating polyphonic music aligned with musical rules. Jiang et al. \cite{jiang2020rl} presented an interactive RL-Duet framework for real-time human-machine duet improvisation. The actor-critic
with generalised advantage estimator (GAE) \cite{schulman2015high} based music generation agent was able to learn a policy to generate musical note based on the previous context. They trained the model on monophonic and polyphonic data and were able to generate high-quality musical pieces compared to a baseline method. Jaques et al.~\cite{jaques2016generating} utilised a deep Q-learning agent with a reward function based on rules of music theory and probabilistic outputs of an RNN. They showed that the proposed model can learn composition rules while maintaining the important information of data learned from supervised training. For audio-based generative models, it is often important to tune the generated samples towards some domain-specific metrics. To achieve this, Guimaraes et al.~\cite{guimaraes2017objective} proposed a method that combines  adversarial training with RL. Specifically, they extend the training process of a GAN framework to include the domain-specific objectives in addition to the discriminator reward. Experimental results show that the proposed model can generate music while maintaining the information originally learned from data, and attained	 improvement in the desired metrics. In~\cite{lee2017polyphonic}, they also used a GAN-based model for music generation and explored optimisation via RL. RaveForce~\cite{lan2019raveforce} is a DRL-based environment for music generation, which can be used to search new synthesis parameters for a specific timbre of an electronic musical note or loop. 

Score following is the process of tracking a musical performance for a known symbolic representation (a score). In~\cite{dorfer2018learning}, the authors modelled the score following task with DRL algorithms such as synchronous advantage actor-critic (A2C). They designed a multi-modal RL agent that listens to music, reads the score from an image, and follows the audio in an end-to-end fashion. Experiments on monophonic and polyphonic piano music showed promising results compared to state-of-the-art methods. The score following task is  studied in~\cite{henkel2019score} using the A2C and proximal policy optimisation (PPO). This study showed that the proposed approach could be applied to track real piano recordings of human performances. 

\begin{figure*}[!h]
\centering
\includegraphics[width=0.8\textwidth]{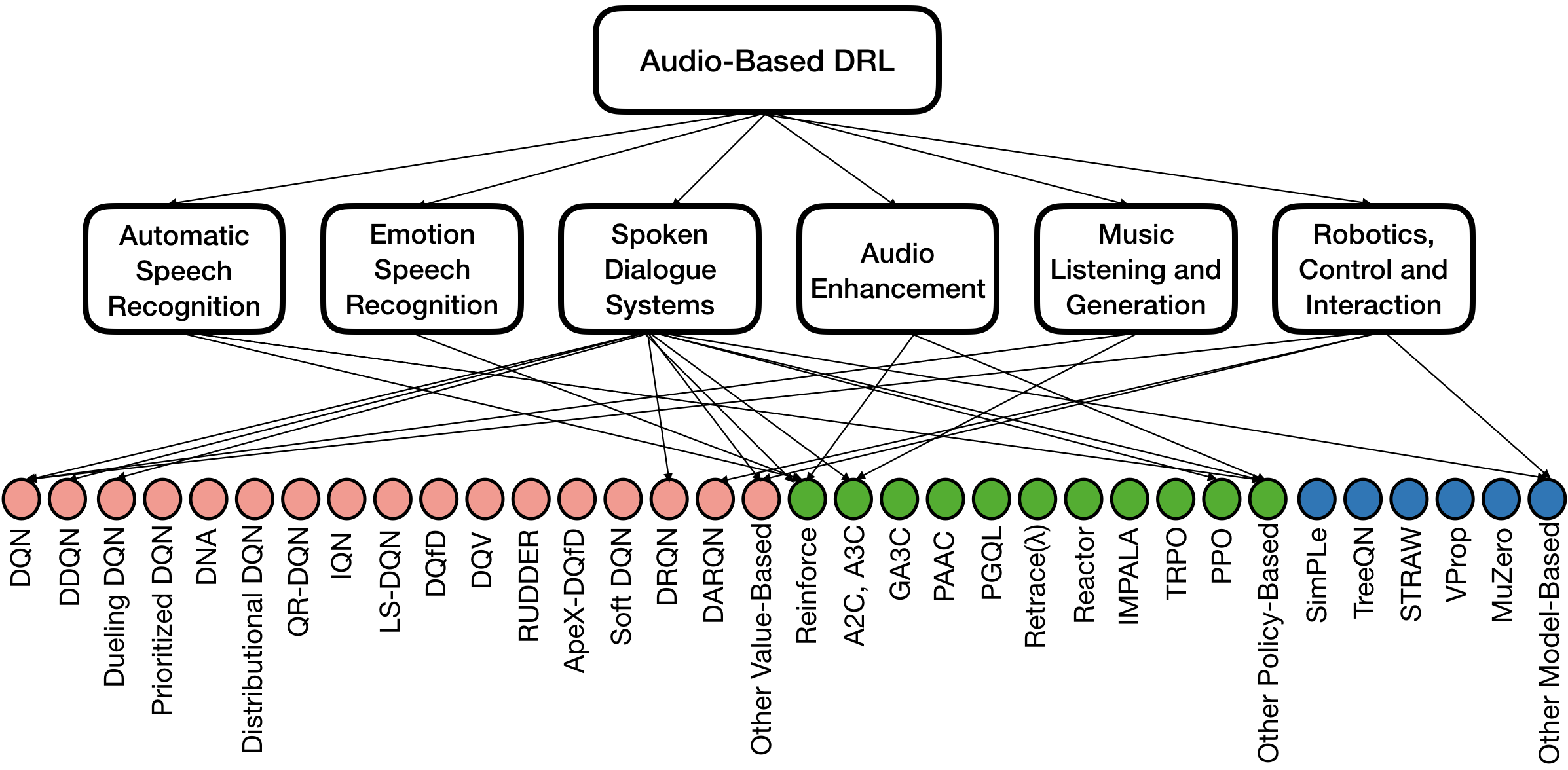}
\caption{A summary of audio-based DRL connecting the application areas and algorithms described in the previous two sections -- the coloured circles correspond to the three groups of algorithms (from left to right: value-based, policy-based, model-based)}
\label{fig:DRL_Apps_Tree}
\end{figure*}

 \subsection{Robotics, Control, and Interaction}
%In recent years, 
There is a recent growing research interest in robotics to enable robots with abilities such as recognition of users' gestures and intentions~\cite{howint}, and generation of socially appropriate speech-based behaviours~\cite{goodrich2007human}. In such applications, RL is suitable because robots are required to learn from rewards obtained from their actions. Different studies have explored different DRL-based approaches for audio and speech processing in robotics. Gao et al.~\cite{gao2020spoken} simulated an experiment for the acquisition of spoken-language to provide a proof-of-concept of Skinner's idea~\cite{skinner1957verbal}, which states that children acquire language based on behaviourist reinforcement principles by associating words with meanings. Based on their results, the authors were able to show that acquiring spoken language is a combination of observing the environment, processing the observation, and grounding the observed inputs with their true meaning through a series of reinforcement attempts. In~\cite{yu2018interactive}, authors build a virtual agent for language learning in a maze-like world. It interactively acquires the teacher's language from question answering sentence-directed navigation. Some other studies~\cite{sinha2019attention,hermann2017grounded,hill2018understanding} in this direction have also explored RL-based methods for spoken language learning.

In human-robot interaction, researchers have used audio-driven DRL for robot gaze control and dialogue management. In~\cite{lathuiliere2019neural}, the authors used Q-learning with DNNs for audio-visual gaze control with the specific goal of finding good policies to control the orientation of a robot head towards groups of people using audio-visual information. Similarly, authors of~\cite{lathuiliere2018deep} used a deep Q-network taking into account visual and
acoustic observations to direct the robot's head towards targets of interest. Based on the results, the authors showed that the proposed framework generates state-of-the-art results. Clark et al.~\cite{clark2018deep} proposed an end-to-end learning framework that can induce generalised and high-level rules of human interactions from structured demonstrations. They empirically show that the proposed model was able to identify both auditory and gestural responses correctly. Another interesting work~\cite{hussain2019speech} utilised a deep Q-network for speech-driven backchannels like laugh generation to enhance engagement in human-robot interaction. Based on their experiments, they found that the proposed method has the potential of training a robot for engaging behaviours. Similarly,~\cite{hussain2019batch} utilised recurrent Q-learning for backchannel generation to engage agents during human-robot interaction. They showed that an agent trained using off-policy RL produces more engagement than an agent trained from imitation learning. 
In a similar strand,~\cite{BuiC19} have applied a deep Q-network to control the speech volume of a humanoid robot in environments with different amounts of noise. In a trial with human subjects, participants rated the proposed DRL-based solution better than fixed-volume robots. DRL has also been applied to  spoken language understanding~\cite{ZamaniMWWF18}, where a deep Q-network receives symbolic representations from an intent recogniser and outputs actions such as \texttt{(keep mug on sink)}.
%Other recent research efforts in DRL-based conversational robots include~\cite{QureshiNYI18,Cuayahuitl20}. 
In~\cite{QureshiNYI18}, the authors trained a humanoid robot to acquire social skills for tracking and greeting people. In their experiments, the robot learnt its human-like behaviour from experiences in a real uncontrolled environment. In~\cite{Cuayahuitl20}, they propose an approach for efficiently training the behaviour of a robot playing games using a very limited amount of demonstration dialogues. Although the learnt multimodal behaviours are not always perfect (due to noisy perceptions), they were reasonable while the trained robot interacted with real human players. Efficient training has also been explored using interactive feedback from human demonstrators as in~\cite{abs-2007-03363}, who show that DRL with interactive feedback leads to faster learning and with fewer mistakes than autonomous DRL (without interactive feedback).

Robotics plays an interesting role in bringing audio-based DRL applications together including all or some of the above. For example, a robot recognising speech and understanding language~\cite{ZamaniMWWF18}, aware of emotions~\cite{lakomkin2018emorl}, carryout activities such as playing games~\cite{Cuayahuitl20}, greeting people~\cite{QureshiNYI18}, or playing music~\cite{abs-2011-05715}, among others. Such a collection of DRL agents are currently trained independently, but we should expect more connectedness between them in the future work.

\section{Challenges in Audio-Based DRL}
\begin{figure}[t]%[!ht]
\centering
\includegraphics[width=0.49\textwidth]{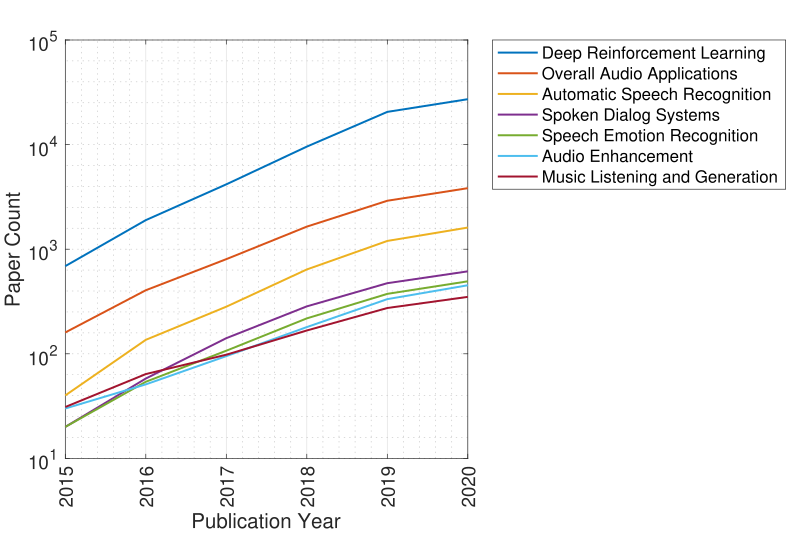}
\caption{Cumulative distribution of publications per year (data gathered from 2015 to 2020) -- from \protect\url{https://www.scopus.com}}
\label{fig:DRLvsaudio}
\end{figure}

The research works in the previous section have focused on a narrow set of DRL algorithms and have ignored the existence of many other algorithms, as can be noted in Figure~\ref{fig:DRL_Apps_Tree}. This suggests the need for a stronger collaboration between core DRL and audio-based DRL, which may be already happening. In Figure~\ref{fig:DRLvsaudio}, we note an increased interest in the communities of core and applied DRL. While core DRL grew from 3 to 4 orders of magnitude from 2015 to 2020, applied DRL grew from 2 to 3 orders of magnitude in the same period.

\begin{figure}[!t]%[!ht]
\centering
\includegraphics[width=0.48\textwidth]{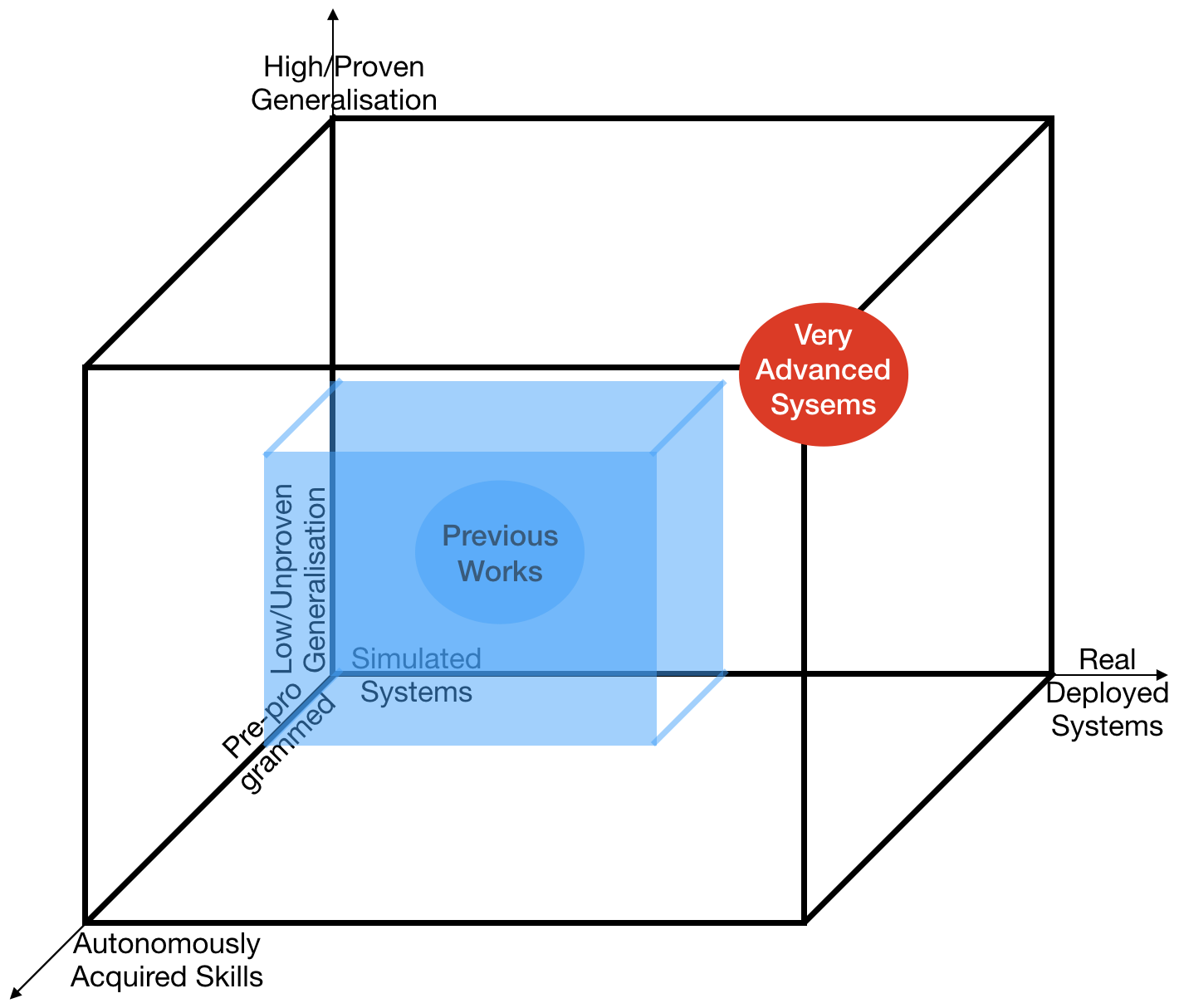}
\caption{A pictorial view of previous works on audio-based DRL and potential dimensions to explore in future systems.}
\label{fig:3D-DRL}
\end{figure}

Figure~\ref{fig:3D-DRL} help us to illustrate that previous works have only explored a modest space of what is possible. Based on the related works above, we have identified three main challenges that need to be addressed by future systems. Those dimensions converge in what we call `very advanced systems'.
%---and challenging indeed.

\label{challenges}
%\subsection{DRL in Audio Based Real Systems}
\subsection{Real-World Audio-Based Systems}
Most of the DRL algorithms described in Section~\ref{sec:DRL} carry out experiments on the Atari benchmark~\cite{BellemareNVB13}, where there is no difference between training and test environments. This is an important limitation in the literature, and it should be taken into account in the development of future DRL algorithms. In contrast, audio-based DRL applications tend to make use of a more explicit separation between training and test environments. While audio-based DRL agents may be trained from offline interactions or simulations, their performance requires to be assessed using a separate set of offline data or real interactions. The latter (often referred to as {\it human evaluations}) is very important for analysing and evidencing the quality of learnt behaviours. 
%Unlike much of the research performed in DRL for audio processing, real systems do not have separate training and evaluation environments. Therefore, DRL agent must perform reasonably well and safely throughout learning. 
In almost all (if not all) audio-based systems, the creation of data is difficult and expensive. This highlights the need for more data-efficient algorithms---specially if DRL agents are expected to learn from real data instead of synthetic data. In high-frequency audio-based control tasks, DRL agents have the requirements of learning fast and avoiding repeating the same mistake. Real-world audio-based systems require algorithms that are sample efficient and performant in their operations. This makes the application of DRL algorithms in real systems very challenging. Some studies such as~\cite{finn2017model,chua2018deep,buckman2018sample}, have presented approaches to improve the sample efficiency of DRL systems. These approaches, however, have not been applied to audio-based systems. This suggests that much more research is required to make DRL more practical and successful for its application in real audio-based systems.    

%%%% This subsection seems to be more general to ML/DL and not specific of audio-based DRL. HC suggests to leave it out %%%%
%\subsection{Robustness of DRL}
%Advances in DNNs has a tremendous impact on the efficiency of RL in audio processing tasks. However, it has recently shown that the DNNs are susceptible to adversarial attacks that mask DNNs into predicting the wrong label by perturbing the input with adversarial noise~\cite{goodfellow2014explaining}. Studies have shown the performance of the state-of-the-art audio processing systems also drop against adversarial attacks~\cite{carlini2018audio,subramanian2020study}. It opens up an exciting frontier regarding the robustness of DNNs in general. Robustness is also critical to enable the successful adoption of DRL in audio processing-related problems. Although, adversarial attacks on DRL agents are different from attacking DNNs, however, studies proved the susceptibility of DRL against adversarial attacks~\cite{lin2017tactics,behzadan2017vulnerability}. Some studies have evaluated various methods to achieve robustness against adversarial attacks in DRL~\cite{pattanaik2018robust}; however, such works also need to be evaluated in audio processing tasks. 

\subsection{Knowledge Transfer and Generalisation}
Learning behaviours from complex signals like speech and audio with DRL requires processing high-dimensional inputs and performing extensive training on a large number of samples to achieve improved performance. The unavailability of large labelled datasets is indeed one of the major obstacles in the area of audio-driven DRL~\cite{purwins2019deep}. Moreover, it is computationally expensive to train a single DRL agent, and there is a need for training multiple DRL agents in order to equip  audio-based systems with a variety of learnt skills. Therefore, some researchers have turned their attention to studying different schemes such as policy distillation~\cite{rusu2015policy}, progressive neural networks~\cite{rusu2016progressive}, multi-domain/multi-task learning~\cite{CuayahuitlYWC17,ultes2017pydial,li2015recurrent,jaderberg2016reinforcement} and others~\cite{yin2017knowledge,nguyen2020deep,glatt2016towards} to promote transfer learning and generalisation in DRL to improve system performance and reduce computational costs. Only a few studies in dialogue systems have started to explore transfer learning in DRL for the speech, audio and dialogue domains~\cite{mo2018personalizing,carrara2018safe,ChenCCTGY18,NarasimhanBJ18,AmmanabroluR19}, and more research is needed in this area. DRL agents are often trained from scratch instead of inheriting useful behaviours from other agents. Research efforts in these directions would contribute towards a more practical,  cost-effective, and robust  application of audio-based DRL agents. On the one hand, to train agents less data-intensively, and on the other to achieve reasonable performance in the real world.

%%%% This subsection reads more like literature review of attention-based DRL rather than a major challenge. HC suggests to move it up  to section III.A (Value-Based DRL) %%%%
%\subsection{Memory and Attention}
%In the audio domain, DNNs with memory units like LSTM has shown great performance, especially for speech modeling. In DRL, studies also shown attempted to incorporate the memory element in DRL algorithms. For instance, deep recurrent Q-network (DRQN) was introduced in~\cite{hausknecht2015deep} was able to successfully integrates information through time and perform well on standard Atari games. Further improvement was made by introducing attention mechanism in DQN, resulting in deep recurrent Q-network (DARQN)~\cite{sorokin2015deep}. This allows DARQN to focus on the specific part of the input and achieve better performance compared to DQN and DRQN on games. Some studies~\cite{oh2016control,parisotto2017neural} also propose methods to incorporate memory in DRL; however, they are not focused on the audio signal. Although attention-based recurrent networks have been evaluated in audio domain  but it will be interesting to design DRL algorithms that can focus on the specific part of the audio signal and can learn contexts from these parts.    

%\subsection{Multi-agent DRL in Audio}
\subsection{Multi-Agent and Truly Autonomous Systems}
 Audio-based DRL has achieved impressive performance in single-agent domains, where the environment stays mostly stationary. But in the case of audio-based systems operating in real-world scenarios, the environments are typically challenging and dynamic. 
For instance, multi-lingual ASR and spoken dialogue systems need to learn policies for different languages and domains. These tasks not only involve a high degree of uncertainty and complicated dynamics but are also characterised by the fact that they are situated in the real physical world, thus have an inherently distributed nature. The problem, thus, falls naturally into the realm of multi-agent RL (MARL), an area of knowledge with a relatively long history, and has recently re-emerged due to advances
in single-agent RL techniques~\cite{littman1994markov, hernandez2019survey}. Coupled with recent advances in DNNs, MARL has been in the limelight for many recent breakthroughs in various domains including control systems, communication networks, economics, etc. However, applications in the audio processing domain are relatively limited due to various challenges. The learning goals in MARL are multidimensional---because the objectives of all agents are not necessarily aligned. This situation can arise for example in simultaneous emotion and speaker voice recognition, where the goal of one agent is to identify emotions and the goal of the other agent is to recognise the speaker. As a consequence, these agents can independently perceive the environment, and act according to their individual objectives (rewards) thus modifying the environment. This can bring up the challenge of dealing with equilibrium points, as well as some additional performance criteria beyond return-optimisation, such as the robustness against potential adversarial agents. As all agents try to improve their policies according to their interests concurrently, therefore the action executed by one agent affects the
goals and objectives of the other agents (e.g. speaker, gender, and emotion identification from speech at the same time), and vice-versa. 
%Furthermore, as the agents adapt and learn their behaviour, the environment faced by each agent becomes non-stationary thus invalidating the basic framework of most theoretical analyses in the single-agent based settings. This further results in suboptimal decision rules locally, as each agent has limited access to the observations of others. Furthermore, the joint action space that increases exponentially with the number of agents may cause scalability issues that can be critical in many audio applications in battery-operated devices.  We believe that specialised targeted work is required, to explore the potential of MARL in the various audio processing tasks, that explicitly address the non-stationary environments to boost performance. 

One remaining challenging aspect is that of autonomous skill acquisition. Most, if not all, DRL agents currently require a substantial amount of pre-programming as opposed to acquiring skills autonomously to enable personalised/extensible behaviour. Such pre-programming includes explicit implementations of states, actions, rewards, and policies. Although substantial progress in different areas has been made, the idea of creating audio-driven DRL agents that autonomously learn their states, actions, and rewards in order to induce useful skills remains to be investigated further. Such kind of agents would have to know when and how to observe their environments, identify a task and input features, induce a set of actions, induce a reward function (from audio, images, or both), and use all of that to train policies. Such agents have the potential to show advanced levels of intelligence, and they would be very useful for applications such as personal assistants or interactive robots.

%SEARCHING THE ABOVE CHALLENGES IN AUDIO.....
 %Most of the work in the audio domain considered single-agent DRL to model in a stationary environment. The environment might become more challenging and dynamic for audio-based systems working in complex situations. For instance, multi-lingual ASR and spoken dialogue system need to learn policies for individual language. In such situations, multi-agent DRL can be exploited, which is challenging in terms of curse-of-dimensionality, communication, and credit assignment~\cite{shao2019survey}.   

%\subsection{Exploration vs. Exploitation}

\section{Summary and Future Pointers}
\label{summary}
This literature review shows that DRL is becoming popular in audio processing and related applications. We collected DRL research papers in six different but related areas: automatic speech recognition (ASR), speech emotion recognition (SER), spoken dialogue systems (SDSs), audio enhancement, audio-driven robotic control, and music generation. 
%\SL{We plot the research studies in these areas and compared with DRL studies in Figure \ref{fig:DRLvsaudio}. Data is gathered from Scopus\footnote{Scopus is a centralised repository for indexing of research literature which index data from all other scholarly repositories} using its database queries for 2015--2020. }
%
%\begin{figure}[t]%[!ht]
%\centering
%\includegraphics[width=0.49\textwidth]{pictures/RL_log_graph.eps}
%\caption{Cumulative distribution of publications per year (data gathered from 2015 to 2020).}
%\label{fig:DRLvsaudio}
%\end{figure}  

\begin{enumerate}
\item In ASR, most of the studies have used policy gradient-based DRL, as it allows learning an optimal policy that maximises the performance objective. We found studies aiming to solve the complexity of ASR models~\cite{dudziak2019shrinkml}, tackle slow convergence issues~\cite{williams1992simple}, and speed up the convergence in DRL~\cite{rajapakshe2020deep}. 
\item The development of SDSs with DRL is gaining interest and different studies have shown very interesting results that have outperformed current state-of-the-art DL approaches~\cite{weisz2018sample}. However, there is still room for improvement regarding the effective and practical training of DRL-based spoken dialogue systems.  
\item Several studies have also applied DRL to emotion recognition and empirically showed that DRL can (i) lower latency while making predictions~\cite{lakomkin2018emorl}, (ii) understand emotional dynamics in communication~\cite{sangeetha2019emotion}, and (iii) enhance human-computer interaction~\cite{chen2017multimodal}. 
\item In the case of audio enhancement, studies have shown the potential of DRL. While these studies have focused their attention on the speech signals, DRL can be used to optimise the audio enhancement module along with performance objectives such as those in ASR~\cite{shen2019reinforcement}. 
\item In music generation, DRL can optimise rules of music theory as validated in different studies~\cite{jaques2016generating,guimaraes2017objective}. It can also be used to search for new tone synthesis parameters~\cite{lan2019raveforce}. Moreover, DRL can be used to perform score following to track a musical performance~\cite{dorfer2018learning}, and it is even suitable for tracking real piano recordings~\cite{henkel2019score}, among other possible tasks. 
\item In robotics, audio-based DRL agents are in their infancy. Previous  studies have trained DRL-based agents using simulations, which have shown that reinforcement principles help agents in the acquisition of spoken language. Some recent works~\cite{hussain2019speech,hussain2019batch} have shown that DRL can be utilised to train gaze controllers and  speech-driven backchannels like laughs in human-robot interaction. 
%In robotics, audio processing using DRL has been used for gaze control. Other studies have trained DRL-based agents using simulations, which have shown that reinforcement principles help agents in the acquisition of spoken language. Some recent works~\cite{hussain2019speech,hussain2019batch} have shown that DRL can be utilised to train gaze controllers and  speech-driven backchannels like laughs in human-robot interaction. 
\end{enumerate}

The related works reviewed above highlight several benefits of using DRL for audio processing and applications. 
%While core DRL (see Section~\ref{sec:DRL}) seems to be ahead of applied DRL (see Section~\ref{survey}) regarding theoretical advances, core DRL looks behind applied DRL with regard to evaluations. Most agents in core DRL have been evaluated only in simulated environments or using datasets recorded in controlled situations. In contrast, most agents in applied DRL have been evaluated using separate environments for training and testing -- including evaluations in realistic environments (and human participants). This suggests a need for closer collaborations between core and applied DRL, which would help designing agents with better understanding of their surroundings and with high levels of generalisation. 
%%future efforts should consider evaluations in real enviroments in order to cope with more complex and ever-changing than simulations. 
%%Addressing this aspect would help designing systems with better understating of their surroundings, enabling them to learn high-level causal relationships with high levels of generalisation. 
Challenges remain before such advancements will succeed in the real world, including endowing agents with commonsense knowledge, knowledge transfer, generalisation, and autonomous learning, among others. Such advances need to be demonstrated not only in simulated and stationary environments, but in real and non-stationary one as in real world scenarios. Steady progress, however, is being made in the right direction for designing more adaptive audio-based systems that can be better suited for real-world settings. If such scientific progress keeps growing rapidly, perhaps we are not too far away from AI-based autonomous systems that can listen, process, and understand audio and act in more human-like ways in increasingly complex environments. 

%%There is still a need to evaluate such agents in real scenarios, where the environment is more complex and ever-changing than simulations. In contrast, applied DRL 
%However, we found that theoretical advances in DRL are ahead of applied DRL in the audio domain. Unfortunately, many of those DRL agents have been evaluated only in simulated environments or using datasets recorded in controlled situations. There is still a need to evaluate such agents in real scenarios, where the environment is more complex and ever-changing than simulations. Addressing this aspect would help designing systems with better understating of their surroundings, enabling them to learn high-level causal relationships with high levels of generalisation. Challenges remain before such advancements will succeed in the real world, including endowing agents with commonsense knowledge, knowledge transfer, generalisation, and autonomous learning, among others. Steady progress, however, is being made in the right direction for designing more adaptive audio-based systems that can be better suited for real-world settings. If such scientific progress keeps growing exponentially, perhaps we are not too far away from AI systems that can listen, understand audio, and act in more human-like ways in increasingly complex environments.

\section{Conclusions}
\label{conclu}
In this work, we have focused on presenting a comprehensive review of deep reinforcement learning (DRL) techniques for audio based applications. We reviewed DRL research works in six different audio-related areas including automatic speech recognition (ASR), speech emotion recognition (SER), spoken dialogue systems (SDSs), audio enhancement, audio-driven robotic control, and music generation. In all of these areas, the use of DRL techniques is becoming increasingly popular, and ongoing research on this topic has explored many DRL algorithms with encouraging results for audio-related applications. Apart from providing a detailed review, we have also highlighted (i) various challenges that hinder DRL research in audio applications and (ii) various avenues for future research. We hope that this paper will help researchers and practitioners interested in exploring and solving problems in the audio domain using DRL techniques.

\section{Acknowledgements}
We would like to thank Kai Arulkumaran (Imperial College London, United Kingdom) and Dr Soujanya Poria (Singapore University of Technology and Design) for proving feedback on the paper. We also thank Waleed Iqbal (Queen Mary University of London, United Kingdom) for helping with the extraction of DRL related data from Scopus (Figure \ref{fig:DRLvsaudio}).

% \bibliographystyle{IEEEtran}
% \bibliography{Reference}


\end{document}